\documentclass[
11pt, 
english,
singlespacing, 
parskip, 
headsepline, 
longbibliography
]{MastersDoctoralThesis} 

\usepackage{setspace}
\usepackage[utf8]{inputenc} 
\usepackage[T1]{fontenc} 
\usepackage{xcolor}
\usepackage{blindtext}

\usepackage{mathptmx}
\usepackage{bm}
\usepackage{epigraph}

\usepackage[style=ieee,citestyle=numeric-comp]{biblatex}
\AtEveryBibitem{%
  \clearfield{issn} 
  \clearfield{doi} 
  \ifentrytype{online}{}{
    \clearfield{url}
  }
}
\usepackage{titlesec}

\titleformat{\subsection}
  {\normalfont\Large\bfseries}{\thesubsection}{1em}{}
\titleformat{\section}
  {\normalfont\LARGE\bfseries}{\thesection}{1em}{}
\titleformat{\subsubsection}
  {\normalfont\large\bfseries}{\thesubsubsection}{1em}{}

\usepackage{bookmark}
\usepackage{multicol}

\usepackage{amsmath}
\usepackage{amssymb}
\usepackage{mathtools}

\usepackage{cases}
\usepackage{amsmath}
\usepackage{dsfont}
\usepackage{xcolor}
 \usepackage{relsize}
\usepackage{soul}

\usepackage{tikz-cd}

\usepackage[autostyle=true]{csquotes} 

\usepackage{changepage}
\usepackage{graphicx}
\usepackage{moresize}

\newcommand*{\bigchi}{\mbox{\large$\chi$}}
\newcommand{\olsi}[1]{\,\overline{\!{#1}}} 
\newcommand{\Mod}[1]{\,\left(\mathrm{mod}\ #1\right)}

\usepackage{wrapfig}

\thesistitle{Quantum search in a non-Markovian environment} 
\supervisor{Prof. Ujjwal Sen} 
\examiner{} 
\degree{BS-MS Dual Degree Programme
} 
\author{Sheikh Parvez Mandal} 
\addresses{} 

\subject{Physics} 
\university{\href{https://www.iiserpune.ac.in/}{Indian Institute of Science Education and Research, Pune}} 
\department{\href{https://www.iiserpune.ac.in/}{\textsc{Indian Institute of Science Education and Research, Pune}}} 
\group{\href{}{Research Group Name}} 
\faculty{\href{}{Faculty Name}} 

\AtBeginDocument{
\hypersetup{pdftitle=\ttitle} 
\hypersetup{pdfauthor=\authorname} 
\hypersetup{pdfkeywords=\keywordnames} 
}
\usepackage[export]{adjustbox}

\makeatletter
\newenvironment{smalleralign}[1][\small]
 {\par\nopagebreak\leavevmode\vspace*{-\baselineskip}%
  \skip0=\abovedisplayskip
  #1%
  \def\maketag@@@##1{\hbox{\m@th\normalfont\normalsize##1}}%
  \abovedisplayskip=\skip0
  \align}
 {\endalign\ignorespacesafterend}
\makeatother

\begin{document}
\def\citepunct{,}

\frontmatter 

\pagestyle{plain} 
\definecolor{graycolor}{rgb}{0.7421875,0.7421875,0.7421875}


\begin{titlepage}
\begin{center}

\HRule \\[0.4cm] 
{\huge \bfseries \ttitle\par}\vspace{0.4cm} 
\HRule \\[.7cm] 
\textsc{\Large A Thesis}\\[0.5cm] 


\Large {submitted to \\[.2cm]
{\Large \deptname}\\[.2cm]
in partial fulfillment of the requirements\\ for the \degreename}\\[0.5cm] 
\textit{by}\\[.5cm]
{\Large\authorname\\[1cm]
\textit{under the supervision of}\\
\supname\\ Harish-Chandra Research Institute,\\ Prayagraj 211019, India}\\
\vfill
 \includegraphics[width=0.15\textwidth]{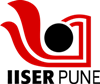} 
 \\[.2cm]
 {\Large Indian Institute of Science Education and Research}\\
{\Large Dr.\ Homi Bhabha Road,}\\
{\Large Pashan, Pune 411008, India}\\[.4cm]
{\Large 2022}
 
\end{center}
\end{titlepage}

\cleardoublepage



\large\noindent


{\renewcommand{\baselinestretch}{1.2}\large
\tableofcontents}

{\renewcommand{\baselinestretch}{0.8}
\listoffigures} 

\vspace{1.5cm}
{\Huge{\textbf{List of symbols and abbreviations}}}
\vspace{.4cm}
\thispagestyle{plain}
\begin{adjustwidth}{.7cm}{}
\setlength{\columnsep}{2cm}
\begin{multicols}{2}
\begin{tabbing}
aaaaaa \=b \=dd \=cc  \kill
\textbf{GSA} \> \>\textbf{G}rover's \textbf{S}earch \textbf{A}lgorithm, \\
\textbf{Qubit} \> \> \textbf{Qu}antum \textbf{bit}; a 2-level system,\\
$\quad \bm{\forall}$ \> \> for all,
\end{tabbing}
\columnbreak
\vfill

\begin{tabbing}
aaaaaa \=bb \=cc  \kill
\textbf{Tr} \>\textbf{Tr}ace, \\
\textbf{CP} \>\textbf{C}ompletely \textbf{P}ositive,\\
$\ \,\bm{\exists}$ \> there exists.
\end{tabbing}
\end{multicols}
\end{adjustwidth}
\thispagestyle{plain}



\begin{abstract}
\addchaptertocentry{\abstractname} 
This thesis explores the effects and origins of a `noise with memory' in the dynamics of an open quantum system.\ The system considered here is a multi-qubit register performing the Grover's quantum search algorithm.\ We show that a Markovian-correlated noise can enhance the efficiency of the algorithm over a time-correlation-less noise.\ We also analytically find the set of necessary and sufficient conditions for the algorithm's success probability to remain invariant with respect to the number of noisy sites in the register and point out that these conditions hold irrespective of the presence of time-correlations in the noise.\ We then investigate the origins of the type of noise considered.\ In this regard, a `collisional model' is constructed that exactly reproduces the noisy evolution of the open system.\ Non-Markovianity in the system's evolution is then assessed using two well-known measures and they are shown to be non-coincident.\ Our model is then slightly modified to accommodate an elementary thermal bath.\ There we show that increasing the bath's temperature increases information drainage from the system.
\end{abstract}











\mainmatter 

\pagestyle{thesis} 

\vspace*{30mm}
\vspace{5cm}
\thispagestyle{empty}
\begin{center}
\begin{tikzpicture}
\node[align=center,draw,thick,minimum width=.8\textwidth,inner sep=9mm] (titlebox)%
{\fontshape{sc}\bfseries{\huge \fontsize{25}{35} Introduction}};
\node[fill=white] (W) at (titlebox.north) {\bfseries \HUGE \color{gray}0};
\node (feat) at ([yshift=9mm]titlebox.north) {\textsc{\huge Part}};
\end{tikzpicture}
\end{center}
\addcontentsline{toc}{part}{Part-0: Introduction}
\vspace{5cm}
\epigraph{``The problems are solved, not by giving new information, but by arranging what we have known since long.''}{Ludwig Wittgenstein (1889-1951)\\\textit{Philosophical Investigations }\textsuperscript{\cite{wittgenstein}}}
\bookmarksetup{startatroot}


\thispagestyle{plain}
\section{Background}
The last few decades saw the advent and flourishing of
the field of quantum information and computation.\ One of the most important classes of discoveries made in this field has to be that of quantum algorithms which provide substantial computational advantages over their classical counterparts.\ The most significant ones include 
the Deutsch-Jozsa algorithm \cite{Deutsch,Deutsch_1}, Shor’s factoring algorithm~\cite{Shor,Shor1}, the quantum search algorithms~\cite{Grover,Grover_2,Boyer,Lidar,Shenvi, Kempe} and the quantum simulation algorithms~\cite{manin, Feynman,lloyd,bernien,zhan}.\ 
The advantages of these quantum algorithms are assumed to be derived from the efficient use of quantum coherence and entanglement.\ After Grover's seminal proposal \cite{Grover,Grover_2} of his quantum search algorithm, which has been shown to be a special case of the more general \textit{amplitude amplification} algorithm~\cite{Brassard,amplitude}, an extensive amount of research effort has been directed towards implementing and studying the effects of noise on the efficiency of the algorithm in an actual quantum device.\ The experimental implementation  of the algorithm was first done using nuclear magnetic resonance techniques~\cite{Chuang_NMR}.\ Later on, the efficiency of the Grover's algorithm was studied in~\cite{Zalka} and a generalization of the algorithm for an arbitrary amplitude distribution was done in~\cite{Lidar}.\ For more works on the quantum search algorithm, see~\cite{abram,GuiLu, Kwiat,Sheng,Long,Biham3,Heinrich,Roland,Xiao} and for experimental implementations, see~\cite{Jones,Vander,Ermakov,Bhattacharya,Zhang, WaltherP,Brickman,DiCarlo,Figg}.\

\subsubsection{\textit{Grover's search on a noisy register}}
Although theoretically more efficient in comparison to its classical counterpart, an actual implementation of a quantum algorithm critically depends on the error-proof fabrication of the relevant quantum register.\ Therefore, studies on the effect of such distortions from the ideal situation, caused by decoherence and noise is important to assess the usefulness and applicability of an algorithm.\ See e.g.~\cite{Bernstein,Preskill,N&C,Barnes}.\
The effect of noise on the Grover's search algorithm
was studied in~\cite{Altaba}, which investigated 
the effect of random Gaussian noise on the algorithm's efficiency at each step.\
A perturbative method was used 
in~\cite{Azuma} to study decoherence in a noisy Grover algorithm where each qubit suffers phase-flip error independently after each step.\ 
The effect of a noisy oracle was considered in~\cite{Tu, Bae}.\ The effect of unitary noise was considered 
in~\cite{Biham} using a noisy Hadamard gate, with unbiased and isotropic noise, uncorrelated in each iteration of the Grover operators.\ An upper bound on the strength of the noise parameters up to which the algorithm works efficiently was deduced.\ 
A comparison of the effects of several completely positive trace preserving maps in the Kraus form on the efficiency and computational complexity of the algorithm was described in~\cite{Gawron}.\ The performance of the algorithm under localized dephasing was studied in~\cite{Reitzner}.\ For more discussions and further ramifications of the effect of noise on the Grover search algorithm, see~\cite{Salas,  Hasegawa, Cohn}.

\thispagestyle{plain}

\subsubsection{\textit{Non-Markovianity and memory}}
Although the Markovian treatment of open systems has been immensely successful in explaining many physical situations, it is also found that often the quantum processes at hand do not satisfy the strict conditions of Markovianity, such as the Born-Markov or the weak coupling approximation.\ With the advent of the field of quantum information and computation, quantum processors need to be made that probably require tightly-packed qubits or a long coherence time.\ These features will lead to spatio-temporal correlations in the noise due to coupling to some environmental degrees of freedom, since making a quantum device completely devoid of noise is almost unrealistic.\ A qunatum process that do not satisfy Markovianity is called non-Markovian.\ Although several approaches \cite{Guo} to define the boundary between Markovianity and non-Markovianity exist, two major directions are based on - one, CP-divisibility of the map (e.g., work by Rivas et al. \cite{RHP}), and two, information flow in and out of the system (e.g., work by Breuer et al. \cite{Breuer_Laine}).\ We will contrast these two measures in assessing the non-Markovianity in our model.
\subsubsection{\textit{Objectives of the project}}
The main objectives are as follows:
\begin{itemize}
    \item To investigate the effect of noise with memory on the efficiency of a quantum algorithm such as the Grover's quantum search.
    \item To detect non-Markovianity in our noise model by following information dynamics and CP-divisiblity of the system dynamics, and, to find out how they are affected by thermal effects.
\end{itemize}

\thispagestyle{plain}

\section{Outline}
The thesis is organised in three Parts.

\textbf{Part-I} reviews the basic concepts to be used through-out the thesis. This part is divided into two chapters.
\medskip\\
In Chapter \ref{Chapter2}, the notion of open quantum systems is introduced along with methods to describe their evolution.\ Some measures of (non-)Markovianity are presented.\ The concept of a quantum channel with memory is introduced along with an example.
\medskip\\
In Chapter \ref{Chapter3}, we briefly describe how some existing quantum algorithms can surpass classical ones and then introduce the Grover's quantum search algorithm along with a framework for analyzing its success probability.

\textbf{Part-II} analyzes the situation when the register performing Grover's algorithm is an open system. This part is divided into two chapters.
\medskip\\
In Chapter \ref{Chapter4}, we introduce our model 
of noise and analytically find the unitaries representing ``good'' noise, i.e., the noises for which the algorithm's success becomes invariant with respect to the number of noise sites.\ The effects of a memory-less noise and a Markovian-correlated noise are then compared, showing that memory in noise may improve the algorithm's efficiency.
\medskip\\
In Chapter \ref{Chapter5}, a `collisional model' is introduced that exactly reproduces the time evolution of our noisy system.\ We show that back-flow of information from the environment into the system happens for a subspace of all parameter values, but the process still remains non-Markovian for most parameter values, even when the back-flow is absent.\ We then introduce an elementary model of a thermal bath and find that increasing temperature leads to increasing information drainage, i.e., decreasing non-Markovianity of the process.
\thispagestyle{plain}

\textbf{Part-III} concludes the thesis with a summary of the outcomes of the project.

\section{Publication}
\textbf{S.\ P.\ Mandal}, A.\ Ghoshal, C.\ Srivastava, and U.\ Sen, “Invariance of success probability in Grover's quantum search under local noise with memory,” (2023),  \\\href{https://doi.org/10.1103/PhysRevA.107.022427}{\textbf{https://doi.org/10.1103/PhysRevA.107.022427}}.\thispagestyle{plain}

\makeatletter\@openrightfalse
\vspace*{30mm}
\vspace{3cm}
\thispagestyle{empty}
\begin{center}
\begin{tikzpicture}
\node[align=center,draw,thick,minimum width=.8\textwidth,inner sep=9mm] (titlebox)%
{\fontshape{sc}\bfseries{\huge \fontsize{25}{35} Basic concepts}};
\node[fill=white] (W) at (titlebox.north) {\bfseries \HUGE \color{gray}I};
\node (feat) at ([yshift=9mm]titlebox.north) {\textsc{\huge Part}};
\end{tikzpicture}
\vspace{.5cm}\\
\textit{This part discusses the main methods and concepts used in the thesis.\\ In Chapter-1, the methods of dealing with the dynamics of open quantum systems are reviewed. \\ In Chapter-2, a brief introduction to 
some of the variants of quantum search algorithms is presented which is followed by a discussion on the Grover's algorithm.}
\end{center}
\addcontentsline{toc}{part}{Part-I: Basic concepts}
\vspace{5cm}
\epigraph{``The main character of any living system is openness.''}{Ilya Prigogine (1917-2003)\\\textit{Nobel Laureate in Chemistry}}\clearpage
\bookmarksetup{startatroot}

\chapter{Open quantum systems} 

\label{Chapter2}
\textit{In this chapter we review the concept of open systems, dynamical maps, master equations and some measures of non-Markovianity.\ We then elucidate the concept of `information' in quantum context.\ The chapter ends with an introduction to quantum channels with memory and a Markovian-correlated noise channel.\ These ideas will be implemented extensively throughout the thesis.}
\section{What is an open system?}
\begin{wrapfigure}[13]{r}{0.37\textwidth}
    \centering
    \includegraphics[width=.99\linewidth]{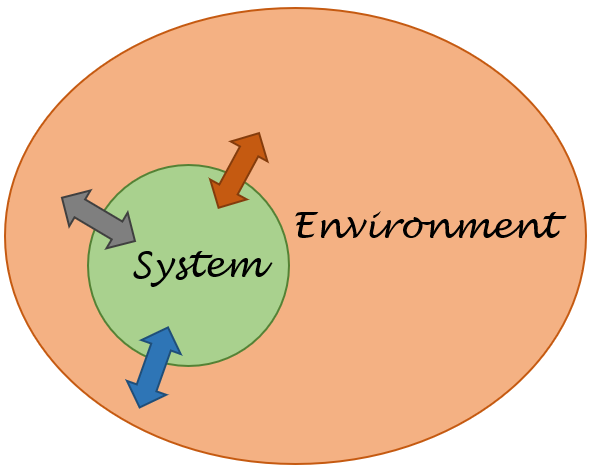}
    \caption[Schematic diagram of an open System]{Schematic diagram of an open System (green) interacting with an Environment (brown); leading to an exchange of energy or information.}
    \label{fig:Open_sys}
\end{wrapfigure}A system coupled to some external degrees of freedom is called an open system \cite{Petru,Riv,Weiss,N&C}.\ The concept, being so general, has found its applications in various studies outside physics, such as in understanding growth of living organisms \cite{Zotin} or in evolutionary theory and in social sciences \cite{Luhmann}.\ In physics, the parts of the Universe $\mathfrak{U}$ that are not included in the \textit{system} $\mathfrak{S}$, could be taken as the \textit{environment} $\mathfrak{E}$. 
But it often turns out that considering a much smaller part of the Universe, $\mathfrak{A} \in \mathfrak{U}$ is enough to determine the dynamics of the open system $\mathfrak{S} \in \mathfrak{A}$, as long as $\mathfrak{A}$ itself is a \textit{closed} system obeying Hamiltonian dynamics.\\ 
If a previously closed system $\mathfrak{S}$ obeying unitary quantum evolution becomes coupled to an environment $\mathfrak{E}$, the \textit{reduced evolution} of $\mathfrak{S}$ no longer remains unitary, in general.\ Any physical quantum system can hardly ever be expected to be perfectly closed.\ As a result, its state can seldom be taken as a \textit{pure state} and a proper description is given by the \textit{density matrix formalism}, first introduced in 1927 independently by Lev Landau \cite{Landau} and von Neumann \cite{Neumann}.


\section{Density matrix formalism}

\subsection{Pure states}
In von Neumann's mathematical formalization of quantum mechanics, a state of a quantum system is a unit vector $|\psi\rangle$ (called \textit{ket} in Dirac notation) which belongs to the \textit{state space} $\mathcal{H}$, which itself is a \textit{Hilbert space} (a complex inner product space that is also a complete metric space \cite{Rudin}.)\ A linear functional (called \textit{bra} in Dirac notation) $\langle \psi |\in \mathcal{H}^*$ gives a number after acting on a ket state.\ The operators are linear functionals from $\mathcal{H}$ to $\mathcal{H}$.\ A physical \textit{observable} is an operator $A$ so that $A = A^{\dagger}$.\ The expectation value of an operator $A$ is $\langle A \rangle = \langle \psi|A|\psi \rangle$.\ One of the most important observables is the energy operator or the Hamiltonian.\\
The time evolution of a pure state $|\psi\rangle$ in \textit{Schr\"odinger picture} under a Hamiltonian $H$ is given as 
\begin{equation}
    |\psi (t)\rangle = e^{-i\, H t} |\psi (0)\rangle,
    \label{eq:Sch_pic}
\end{equation}
where $t$ is time and the reduced Plank's constant, $\hslash$, is set to 1 here and for the remainder of this work.\ The \textit{unitary time-evolution} operator on $\mathcal{H}$ is defined as $U(t,t_0) = e^{-i\,H(t-t_0)}$,  $U(t,t_0)^{\dagger}U(t,t_0)=\mathds{1}$, with $\mathds{1}$ being the identity operator on $\mathcal{H}$.\ So, Eq.\ \eqref{eq:Sch_pic} can also be written as $|\psi (t)\rangle$ $= U(t,0)$ $|\psi (0)\rangle$.\ For a closed system evolving under a time-dependent hamiltonian $H(t)$, the time-evolution operator is given as 
\begin{equation}
    U(t,t_0) = \mathcal{T}e^{-i\, \int_{t_0}^{t}ds\, H(s)}
\end{equation}
where $\mathcal{T}$ is the time-ordering operator.\\ 
In the alternative \textit{Heisenberg picture}, the operators evolve in time instead of the system's state. The evolution of an observable $A$ is given as
\begin{equation}
    A(t) = U(t,t_0)^{\dagger} A(t_0) U(t,t_0).
\end{equation}

\subsection{Mixed states}
Often it is not possible to write down the state of a system as a simple pure state, for example, when the system is a convex mixture of pure states $\{|\psi_{i}\rangle\}$, each occurring with probabilities $\{p_i\}$.\ Then, the state is given by a \textit{density matrix}
\begin{equation}
    \rho = \sum_{i} p_i |\psi_{i}\rangle\langle \psi_{i}|.
\end{equation}
So, $\rho$ is positive semi-definite and $\text{Tr}\{\rho\}=1$.\ For a pure state, $\rho = |\psi\rangle\langle \psi| \implies \text{Tr}\{\rho^2\} = 1$. Whereas for a mixed state, $\text{Tr}\{\rho^2\} < 1$.\ Expectation value of an operator $A$ for the state $\rho$ is $\langle A\rangle = \text{Tr}\{A\rho\}$.\ The space that contains $\rho$ is called the \textit{Liouville space} and the time evolution of $\rho$ is determined by the Liouville-von Neumann equation,
\begin{equation}
    \partial_t \rho = -i[H,\rho] \implies \rho(t) = U(t,t_0)\rho(t_0) U^{\dagger}(t,t_0).
    \label{eq:vN_eq}
\end{equation}

\section{Time evolution of an open system}
 The interactions \cite{Weiss} with the environment generally leads to deviation from unitary dynamics and the system is said to be evolving under \textit{quantum noise} \cite{Gardiner}. It turns out that after tracing out the environment $E$ in the Liouville-von Neumann equation \eqref{eq:vN_eq} for the whole system $\rho_{tot}$, we can write for the open system $\mathfrak{S}$ density matrix $\rho$,
 \begin{gather}
      \partial_t \rho = \mathcal{L}(t)\rho,
      \label{eq:Lindblad_0}\\
      \text{or},\ \ \rho(t) = \mathcal{T}e^{ \int_{t_0}^{t}ds\,\mathcal{L} (s)}\rho(t_0),\\
      \qquad = \Phi(t,t_0)\rho(t_0) \label{eq:Phi}
 \end{gather}
  where $\mathcal{L}(t)$ is called the \textit{lindbladian} (or, quantum \textit{liouvillian}) \cite{Petru} and is the generator for $\Phi(t,t_0)$, a superoperator representing the \textit{dynamical map} (also called \textit{quantum channel}) acting on $\rho$.
 
 \subsection{The dynamical map}
 Thus the dynamical map $\Phi(t,t_0)$ in \eqref{eq:Phi} maps the initial state of the system at time $t_0$ to the state at a time $t$.\ 
\[\begin{tikzcd}[column sep=7em]
\rho_{tot}(0) = \rho(0)\otimes \rho_E \arrow{r}{\text{evolution under}\ U} \arrow[swap]{d}{\text{Tr}_E} & \rho_{tot}(t) = U(t,0)(\rho(0)\otimes \rho_E)U(t,0)^\dagger \arrow{d}{\text{Tr}_E} \\
\rho(0) \arrow{r}{\text{dynamical map}\ \Phi} & \rho(t)= \Phi(t,0)\rho(0)
\end{tikzcd}
\]

\subsection{Kraus-Sudarshan representation}\begin{wrapfigure}[12]{l}{0.32\textwidth}
    \centering
    \includegraphics[width=.99\linewidth]{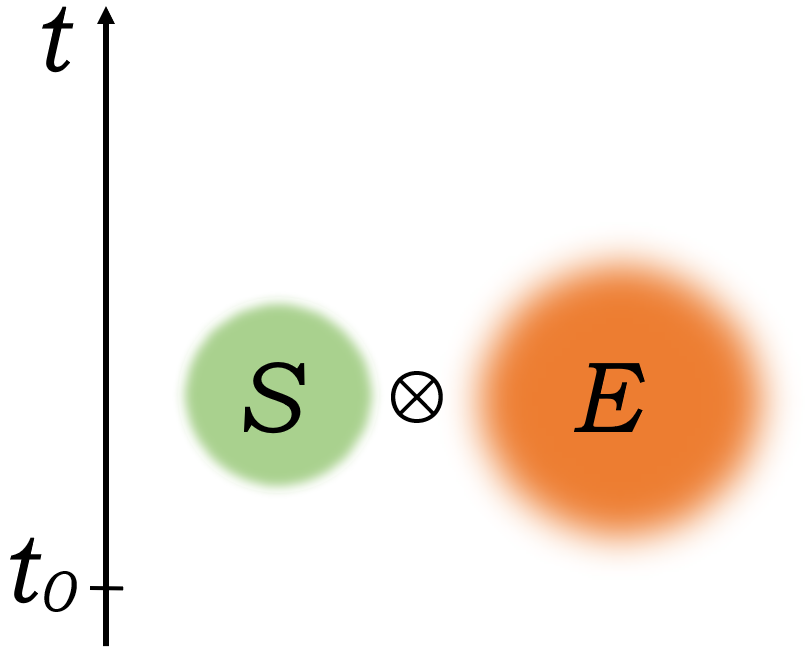}
    \caption[Initial product state of the total system]{The system $S$ and environment $E$ are initially in a product state.}
\end{wrapfigure}

If we decompose the environment's initial state $\rho_E$ as a convex mixture $\{\lambda_{\alpha}\}$ of pure states $\{|e_{\alpha}\rangle \}$,
\begin{equation}
    \rho_E = \sum_{\alpha}\lambda_\alpha |e_\alpha\rangle\langle e_\alpha|,
    \label{eq:e_dec}
\end{equation}
and insert Eq.\ \eqref{eq:e_dec} into the following expression for unitary evolution of an \textit{initially product} joint state of the system and the environment
\begin{equation}
    \rho_{tot}(t) = U(t,0)(\rho(0)\otimes \rho_E)U(t,0)^\dagger,
    \label{eq:sep_rho}
\end{equation}
we get the evolved state of the system after time $t$
\begin{equation}
    \rho(t) = \text{Tr}_E\{\rho_{tot}(t)\} = \sum_{\alpha \beta}^{}K_{\alpha \beta}(t)\rho(0) K_{\alpha \beta}^{\dagger}(t) 
    \label{eq:Op_sum}
\end{equation}
  where $K_{\alpha\beta}(t) = \sqrt{\lambda_{\beta}}\langle e_{\alpha}|U(t,0)|e_{\beta}\rangle$ are the \textit{Kraus operators} and they satisfy the \textit{completeness relation} 
  \begin{equation}
      \sum_{\alpha \beta}^{}K_{\alpha \beta}^{\dagger}(t)K_{\alpha \beta}(t) = \mathds{1},
      \label{eq:Op_sum_1}
  \end{equation}
  which guarantees that $\text{Tr}_{\mathfrak{S}}\{\rho(t)\}=1$. The representation Eq.\ \eqref{eq:Op_sum} is known as the \textit{operator-sum or Kraus-Sudarshan representation} \cite{Sudarshan,Kraus,N&C}.\\
  The dynamical map with the representation Eqs.\ \eqref{eq:Op_sum} and \eqref{eq:Op_sum_1} is a \textit{completely positive (CP) \footnote{A map $\Phi: A\rightarrow B$ is \textit{\textbf{completely positive}} (CP) if the map $(\mathds{1}_{\mathbb{C}^{m\times m}}\otimes \Phi)$ is positive\textsuperscript{2} $\forall\ m$. CP'ty guarantees that the map $\Phi$ maps the set of physically possible density matrices (positive operators) to itself even when the system is coupled to an ancilla. }\textsuperscript{,}\footnote{A map $\Phi: A\rightarrow B$ is \textit{\textbf{positive}} (P) if it maps positive-semidefinite operators ($P\geq0,\ P\in A$) to positive-semidefinite operators ($Q\geq0,\ Q\in B$).}, trace preserving (TP)} map according to the Kraus representation theorem \cite{Choi}. 
  \subsubsection{\textit{Unitality}}
  A \textit{unital map} $\Phi$ takes the maximally mixed $d$-dimensional state $\rho_{mm} = \frac{\mathds{1}}{d}$ to the same state $\Phi(\rho_{mm}) = \frac{\mathds{1}}{d}$. This implies, the Kraus operators, $K_\mu$, for the map, $\Phi$, need to satisfy $\sum_{\mu}^{}K_{\mu}(t)K_{\mu}^{\dagger}(t) = \mathds{1}$.
  A map that is not identity preserving, so that $\Phi(\frac{\mathds{1}}{d}) \neq \frac{\mathds{1}}{d}$, is called \textit{non-unital}.\\
  Unital maps describe \textit{diffusion} or decoherence. Non-unital maps describe \textit{dissipative} processes and can be related to the classical idea of dissipation as the contraction of phase space volume \cite{mata}.

\subsection{(Non-)Markovian continuous time evolution}
We will denote the set of positive (P) dynamical maps from a liouville space to itself as $\mathfrak{P}$, of CP maps as $\mathfrak{P}^{+}$, of trace-preserving (TP) P maps as $\mathfrak{T}$ and of TP CP maps as $\mathfrak{T}^{+}$.
\subsubsection{\textit{Divisibility of a map}}
  Let us consider a map $\Phi \in \{\mathfrak{T},\mathfrak{T}^{+}\}$.\ If there exists a decomposition $\Phi=\Phi_1 \circ \Phi_2$ so that none of $\Phi_i$ is an unitary operator, then $\Phi$ is called a \textit{divisible map} \cite{Cirac}.\\
  Otherwise,
  it is called an \textit{indivisible map}. We will denote the set of divisible maps as $\mathfrak{D}$. 
  
  \subsubsection{\textit{CP- and P-divisibility}}
  If an invertible $\Phi(t,t_0)\in \mathfrak{P}^{+}$ can be decomposed as
  \begin{equation}
        \Phi(t,t_0)=\Phi(t,s)\circ \Phi(s,t_0); \quad t\geq s\geq t_0
        \label{eq:CP_divis}
  \end{equation}
 so that $\Phi(t,s)$ is CP- (P-) map, then $\Phi(t,t_0)$ is called \textit{CP- (P-)divisible}.
  
  \subsubsection{\textit{Markovian maps}}
  It is an important special case when a dynamical map satisfies the \textit{homogeneous} composition law
  \begin{equation}
      \Phi(t_2-t_1+t_0,t_0)\circ \Phi(t_1,t_0) = \Phi(t_2,t_0); \quad t_2\geq t_1\geq t_0,
      \label{eq:semigroup}
  \end{equation}
  analogous to the classical Chapman-Kolmogorov equation.\ Then the corresponding dynamics is said to be \textit{Markovian} and \eqref{eq:Lindblad_0} leads to a \textit{master equation in the strict GKSL (Gorini-Kossakowski-Sudarshan-Lindblad)} form \cite{Gorini}:
\begin{gather}
      \partial_t \rho = \mathcal{L}\rho= -i[H,\rho] + \sum_{\nu} \mathcal{D}_{\nu}[\rho]
      \label{eq:Mark_lind}\\
      \mathcal{D}_{\nu}[\rho] 
      = \gamma_{\nu}\left( L_{\nu}\rho L_{\nu}^{\dagger} - \frac{1}{2}\{L_{\nu}^{\dagger}L_{\nu},\rho\}\right)\\
      \qquad \quad = \frac{\gamma_{\nu}}{2}\left( [L_{\nu}\rho,L_{\nu}^{\dagger}]+ [L_{\nu}\rho,L_{\nu}^{\dagger}]^{\dagger}  \right)
      \label{eq:dissipator}
\end{gather}  
  where $\mathcal{D}_{\nu}[\rho]$ is called the \textit{dissipator}, $\{\gamma_{\nu}\}$ are the positive \textit{decay rates} and $\{L_{\nu}\}$ is the set of orthonormal \textit{lindblad operators} \cite{Havel}. The time-\textit{independent} lindbladian $\mathcal{L}$ in \eqref{eq:Mark_lind} generates the one-parameter \textit{semigroup} $\Phi(t-t_0)=e^{(t-t_0)\mathcal{L}},\ \forall t\geq 0$ \cite{Lind}. Denoting the set of markovian maps as $\mathfrak{M}$, we have $\mathfrak{M}  \subset \mathfrak{D}$.\\
  The Markovian case \eqref{eq:semigroup}-\eqref{eq:dissipator} is not usually the situation in practice. The processes in which this treatment does not hold are called \textit{non-markovian} and their equations can be derived directly from the Liouville-von Neumann equation \eqref{eq:vN_eq} using the \textit{projection operator techniques}. 
  
    \subsubsection{\textit{The Nakajima-Zwanzig equation}}

  From \eqref{eq:vN_eq}, we can define the Liouville operator for the whole system dynamics $\mathcal{L}[\cdot] = i[\cdot,H]$ and the projection super-operator $\mathcal{P}$ 
  \begin{equation}
      \rho_{tot} \xmapsto[]{} \mathcal{P}\rho_{tot} = \text{Tr}_E\{\rho_{tot}\}\otimes \rho_E = \rho \otimes \rho_E
  \end{equation}
 where $\rho_E$ is a fixed environment state so that $\mathcal{P}^2 \rho_{tot} = \mathcal{P}\rho_{tot}$. An orthogonal projection operator $\mathcal{Q}$ is defined with $\mathcal{P}+\mathcal{Q}=\mathds{1}$. Operating with $\mathcal{P}$ and $\mathcal{Q}$ on the total density matrix of \eqref{eq:vN_eq}, we get the system of equations
  \begin{smalleralign}[\normalsize]
     \partial_t \mathcal{P}\rho_{tot} &=\mathcal{P}\mathcal{L}(\mathcal{P}+\mathcal{Q})\rho_{tot},\\
    \text{and}\quad \partial_t \mathcal{Q}\rho_{tot} &= \mathcal{Q}\mathcal{L}(\mathcal{P}+\mathcal{Q})\rho_{tot}.
 \end{smalleralign}
 Solving these and assuming uncorrelated initial state $\rho_{tot}(0) = \rho(0)\otimes \rho_E$ from \eqref{eq:sep_rho}, i.e., $\mathcal{P}\rho_{tot}(0)=\rho_{tot}(0),\ \mathcal{Q}\rho_{tot}(0)=0$, we get the final form for the \textit{Nakajima-Zwanzig equation} \cite{Nakajima,Zwanzig,prigogine}:
  \begin{smalleralign}[\normalsize]
    \partial_t \rho &= -i[H_s,\rho(t)]+\int_{0}^{t} d\tau\  \text{Tr}_E\{\mathcal{P}\mathcal{L}\ e^{\mathcal{Q}\mathcal{L}\tau}\mathcal{Q}\mathcal{L}\mathcal{P}\ \rho_{tot}(t-\tau)\}\\
     &=-i[H_s,\rho(t)]+\int_{0}^{t} d\tau\  \text{Tr}_E\{\mathcal{K}(\tau)\ \mathcal{P}\ \rho_{tot}(t-\tau)\}\\
     &=-i[H_s,\rho(t)]+\int_{0}^{t} d\tau\  \text{Tr}_E\{\mathcal{K}(t-\tau)\ \mathcal{P}\ \rho_{tot}(\tau)\}\label{eq:NZ_eqn}\\
   \text{where} \quad  \mathcal{K}(t) &=\mathcal{P}\mathcal{L}\ e^{\mathcal{Q}\mathcal{L}t}\mathcal{Q}\mathcal{L}\mathcal{P}
\end{smalleralign} 
is the \textit{memory kernel}.\ The equation is linear in $\rho$ but is non-local in time. The dynamics can thus account for non-markovianity and memory effects of an environment.
 
 \subsubsection{\textit{Time-dependent GKSL master equation}}
 The time non-local Eq.\ \eqref{eq:NZ_eqn} can be brought to the `canonical' form as the GKSL equation \eqref{eq:Mark_lind} using the \textit{time-convolutionless} projection operator technique \cite{Chaturvedi,Shibata} under the sufficient condition of the existence of left inverse $\Phi(t,t_0)^{-1}$ for all $t$ so that the lindbladian can be written as 
 \begin{gather}
     \mathcal{L}(t) = (\partial_t \Phi(t-t_0))\  \Phi(t,t_0)^{-1} \label{eq:TCL_lind}\\
     \implies \partial_t \rho = \mathcal{L}(t)\ \rho= -i[H(t),\rho] + \sum_{\nu} \mathcal{D}_{\nu}(t)[\rho]
 \end{gather}
 which is the time-local \textit{time-dependent GKSL} master equation with time- dependent dissipators
 \begin{equation}
     \mathcal{D}_{\nu}(t)[\rho] 
      = \gamma_{\nu}(t)\ \left( L_{\nu}(t)\rho L_{\nu}(t)^{\dagger} - \frac{1}{2}\{L_{\nu}(t)^{\dagger}L_{\nu}(t),\rho\}\right).
      \label{eq:td_diss}
 \end{equation}
 From Eq.\ \eqref{eq:TCL_lind}, the dynamical map can be written as 
 \begin{gather}
     \Phi(t,t_0) = \mathcal{T}e^{ \int_{t_0}^{t}ds\,\mathcal{L} (s)}
     \label{eq:28}\\
     \implies \Phi(t,s)\circ \Phi(s,t_0) = \Phi(t,t_0); \quad t\geq s\geq t_0,
     \label{eq:29}
 \end{gather}
 which is a time-\textit{inhomogeneous} \cite{Man_rev} composition law and differs from the time-\textit{homogeneous} Eq.\ \eqref{eq:semigroup} by being explicitly dependent on the intermediate time $s\geq t_0$ as a result of Eq.\ \eqref{eq:28}.\\
 Since the decay rates $\gamma_\nu (t)$ are time-dependent, they can even be negative temporarily without violating CP'ty of the dynamical map. But the dynamics is 
 \begin{equation}
     \mbox{CP-divisible \textit{if and only if} $\gamma_\nu (t) \geq 0,\ \forall t,\nu$\  \cite{Breuer_NM_rev}.}
     \label{CP-div}
 \end{equation}
 In \cite{Chru}, the condition for markovianity is that
 \begin{equation}
     \mbox{\textit{if} $\Phi(t,t_0)$ is CP-divisible, \textit{then} $\Phi(t,t_0)$ is \textit{markovian}.} \label{CP_marko}
 \end{equation}
 If $\Phi(t,t_0)$ is invertible, then the dynamics is 
 \begin{equation}
     \mbox{P-divisible \textit{if and only if} $\Phi(t,t_0)$ is markovian \cite{Kossa_1,Kossa_2}.}
 \end{equation}

\subsubsection{\textit{Measure of non-markovianity: CP-divisibility}}
The following statement is generally made about non-markovianity from the TD GKSL equation \eqref{eq:td_diss}
\begin{equation}
    \mbox{\textit{if} at least one $\gamma_\nu (t) < 0$, \textit{then} $\Phi(t,t_0)$ is \textit{non-markovian} \cite{hierarchy}.}
\end{equation}
Combining this with statement \ref{CP-div} implies
\begin{equation}
    \mbox{\textit{if} $\Phi(t,t_0)$ is \textit{not CP-divisible}, \textit{then} it is non-markovian.}
\end{equation}
Thus, a measure for non-markovianity can be the non-CP divisibility character of the dynamical map. This is the idea behind the \textit{Rivas-Huelga-Plenio (RHP) measure} \cite{RHP} of non-markovianity,
\begin{gather}
    \mathcal{N}_{RHP}(\Phi) = \underset{\epsilon \rightarrow 0^+}{\mathrm{lim}} 
    \int_{t_0}^{\infty}dt\  \frac{\text{Tr}\{\sqrt{Q^{\dagger}Q}\} \ -\ 1}{\epsilon}\\
    Q = (\Phi_{(t+\epsilon,t)}\otimes \mathds{1}_A)(|\Psi\rangle_{SA}\ {}_{SA}\langle \Psi|)
    \label{eq:N_RHP}
\end{gather}
where $|\Psi\rangle_{SA}$ is the maximally entangled state of the system S with an ancilla A and $Q$ is the \textit{Choi matrix} \cite{Choi} corresponding to $\Phi(t+\epsilon,t)$. \\
If $\Phi_{(t+\epsilon,t)}$ is \textit{CP}, we have Choi matrix $Q\geq 0$, i.e., $\text{Tr}\{\sqrt{Q^{\dagger}Q}\}=1$, implying $\mathcal{N}_{RHP}(\Phi)=0$. \\
Whereas if 
$\Phi_{(t+\epsilon,t)}$ is \textit{not CP}, we have $\text{Tr}\{\sqrt{Q^{\dagger}Q}\}>1$ because of the contraction property of trace distance under CP maps and thus $\mathcal{N}_{RHP}(\Phi)>0$.
Thus $\mathcal{N}_{RHP}$ may detect some P-divisible (but not CP-divisible) maps as non-markovian processes \cite{Guo,Hall,eternals}. 

 \subsubsection{\textit{Stronger measure of non-Markovianity: Information flow}}
 \label{sec:N_BLP}
 Since in a Markovian process an open system continuously loses correlations and information to the environment, a non-Markovian process can be associated with a \textit{backflow of information} from the environment and into the system.\ \begin{wrapfigure}[14]{l}{.37\linewidth}
    \centering
    \captionsetup{width=\linewidth}
    \includegraphics[width=\linewidth,frame]{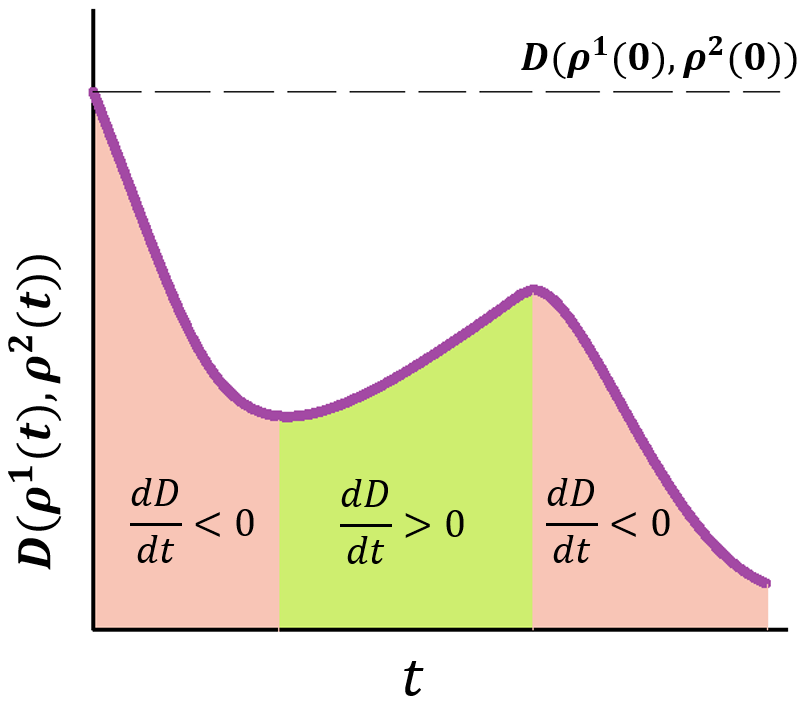}
    \caption[Schematic evolution of trace distance for a CP map]{A schematic evolution of trace distance (purple curve) for a CP map. The initial trace distance between $\rho^1(0)$ and $\rho^2(0)$ is marked as a dashed line and is the upper bound for all $t$.\ But after the initial (red interval, left) decrease, the trace distance increases for some finite time interval (green) before again decreasing (red interval, right).}
    \label{fig:Tr_dist_NM}
\end{wrapfigure} Since two states are more distinguishable when there is more information present in the open system, information flowing in and out of it can be tracked by quantifying their distinguishability and is measured by the \textit{trace distance}
\begin{smalleralign}[\normalsize]
      D(\rho^1,\rho^2)=\frac{1}{2} \text{Tr}\left\{\sqrt{(\rho^1-\rho^2)^{\dagger} (\rho^1-\rho^2)}\right\}
      \label{eq:Tr_dist}
 \end{smalleralign}
 between any two given density matrices $\rho^1$, $\rho^2$ of the open system.\\
 It turns out that under CP dynamical maps, $D(\rho^1(t)$,$\rho^2(t))$ \textit{never exceeds} the initial $D(\rho^1(0)$, $\rho^2(0))$. But, its evolution is not monotonic, for example, see the figure \ref{fig:Tr_dist_NM}.
In the time intervals at which $D(\rho^1(t)$, $\rho^2(t))$ decreases, corresponding to loss of information from the open system to the environment.\ In the Figure \ref{fig:Tr_dist_NM}, there is also a time interval in which the trace distance increases, i.e., $\frac{dD}{dt} >0$, which corresponds to flow of information from the environment into the system. This corresponds to a non-Markovian evolution of the open system and can be quantified by the \textit{Breuer-Laine-Piilo (BLP) measure} \cite{Breuer_Laine} of non-Markovianity, 
\begin{equation}
    \mathcal{N}_{BLP}(\Phi(t,t_0)) = \underset{\rho^1(0),\rho^2(0)}{\mathrm{max}} 
    \int_{ \frac{dD}{dt} >0} dt\  \frac{d}{dt} D(\rho^1(t),\rho^2(t))
    \label{eq:N_BLP}
\end{equation}
  for the time evolution of the open system from $t_0$ to $t$.\ Since for a Markovian dynamical map, $\Phi_{M}\in \mathfrak{M}$, $D(\rho^1(t),\rho^2(t))$ is monotonically decreasing, we have $\mathcal{N}_{BLP}(\Phi_M)=0$.\ For a non-Markovian map $\Phi_{NM}$,\ $\mathcal{N}_{BLP}(\Phi_{NM})>0$.
   If a map is CP-divisible, then there is no information back-flow, but the converse may not be true \cite{Haikka,Guo}.\ Since the growth of trace distance breaks P-divisibility \cite{Breuer_NM_rev}, $\mathcal{N}_{BLP}$ \eqref{eq:N_BLP} quantifying the back-flow of information is considered a measure for \textit{strong} \cite{bernardes} or \textit{essential} non-Markovianity \cite{Chru} as compared to the $\mathcal{N}_{RHP}$ measure \eqref{eq:N_RHP} which quantifies breaking of CP-divisibility.\ There are processes that break CP-divisibility but are P-divisible \cite{Guo,Hall,eternals}.\ $\mathcal{N}_{RHP}$ can detect these processes as non-Markovian while $\mathcal{N}_{BLP}$ can not.

 \subsection{Discrete-time evolution}
 
 Although an open system may evolve continuously in time, it may be useful in some cases to describe its evolution as `stroboscopic' or as discrete time-steps.\ 
 \begin{wrapfigure}[10]{r}{.45\linewidth}
    \centering
    \captionsetup{width=\linewidth}
    \includegraphics[width=\linewidth,frame]{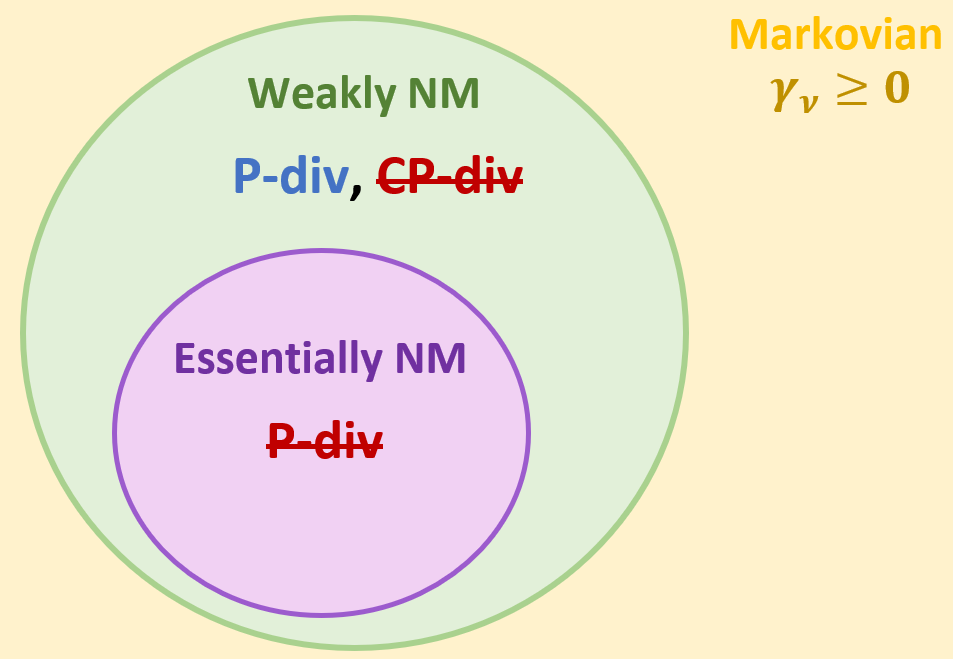}
    \caption[Venn diagram for non-Markovianity]{This is a Venn diagram for \textit{weakly} and \textit{essentially} non-Markovian processes.\ It shows that they break CP- and P- divisibility respectively.}
\end{wrapfigure}Thus, a collection of CPTP maps $\{\Phi_t\}_{t\geq 0}$ can be defined so that the density matrix $\rho_s$ at time $s$ is given as $\rho_s = \Phi_s\, \rho_0$ where $\rho_0$ is the initial state and $\Phi_0=\mathds{1}$.\ 
 
 \subsubsection{\textit{CP-divisibility}}

 Similar to the continuous-time case (Eq.\ \eqref{eq:CP_divis}), the discrete-time dynamical map $\{\Phi_t\}_{t\geq 0}$ is defined to be CP-divisible \textit{if and only if} there exist CP maps $\Phi_{t,s}$ with $t>s>0$, so that $\Phi_t = \Phi_{t,s}\, \Phi_s$.
 \subsubsection{\textit{Information-decreasing maps}}
A discrete-time dynamical map $\{\Phi_t\}_{t\geq 0}$ is information-decreasing \cite{Buscemi,ujan} \textit{if and only if} $\forall$ initial ensemble of states $\{p_i,\rho^i_0\}$, we have 
\begin{equation}
    \underset{\{P^i\}}{\mathrm{\text{max}}}\big( \sum_i p_i\, \text{Tr} \{P^i\,\rho^i_s\}\big)\ \geq\ \underset{\{P^i\}}{\mathrm{\text{max}}}\big( \sum_i p_i\, \text{Tr} \{P^i\,\rho^i_{s+1}\}\big), \quad \forall s\geq 0,
\end{equation}
where the maximization is over all POVM\footnote{A \textit{\textbf{Positive Operator-valued Measure}
} (POVM) is a set of positive semi-definite hermitian matrices $\{P^i\}$ on a Hilbert space so that $\sum_i P^i = \mathds{1}$.}'s $\{P^i\}$'s over the system's Hilbert space.\\
In \cite{Buscemi}, it was shown that a discrete-time dynamical map $\{\Phi_t\}_{t\geq 0}$ is CP-divisible \textit{if and only if}
\begin{equation}
      \mbox{$\{(\mathds{1}_{S'}\otimes (\Phi_t)_{S})\}_{t\geq 0}$ is \textit{information-decreasing} for any ancilla $S'$.} \label{CP_disc}
 \end{equation}
 For an initial ensemble $\{p^1,\rho^1_0;p^2,\rho^2_0\}$, we have, $\mathrm{\text{max}}\big( \sum_i p_i\, \text{Tr} \{P^i\,\rho^i_{s+1}\}\big)$ $= \frac{1}{2}(1+D(p^1\rho^1_0,p^2\rho^2_0))$ \cite{helstrom}.\ Thus, from statement \ref{CP_disc}, we see that\\
 \textit{if} $\{\Phi_t\}_{t\geq 0}$ is CP-divisible, \textit{then} for any hermitian $R$ on $\mathcal{H}_{S'}\otimes\mathcal{H}_S$,
 \begin{equation}
     \mbox{$D\big([\mathds{1}_{S'}\otimes (\Phi_t)_{S}](R),0\big)$ is a non-increasing function of $t$.}
     \label{st_Cp_d}
 \end{equation}
 
 \subsection{Meaning of `Information'}
 Entropy measures how chaotic a system is.\ A system contains more `information' in it means it has lower entropy than a system with less information.\ The \textit{von Neumann entropy} $S$ is the quantum counterpart of the classical Gibbs entropy or the Boltzmann equation.\ For a density matrix $\rho$, it is given as 
 \begin{equation}
     S = -k_B\ \text{Tr}\{\rho\ \text{log}\rho \}.
 \end{equation}
 Thus a system `losing information' to the environment refers to the system's `increasing von Neumann entropy'.\ $S(\rho)=0$ for a pure state $\rho$. Some important properties of $S$ are listed below:
 \begin{itemize}
\item $S(\rho) = S(U\rho U^{\dagger})$ for any unitary transformation $U$.
\item $S(\sum _{i} \lambda_i\, \rho_i) \geq \sum _{i} \lambda_i\ S(\rho_i);\ \sum_i \lambda_i = 1,\ \lambda_i\geq 0$.
\item $S(\text{Tr}_{A}\{\rho\})+S(\text{Tr}_{B}\{\rho\})\geq S(\rho) \geq |S(\text{Tr}_{A}\{\rho\})-S(\text{Tr}_{B}\{\rho\})|$ for a bipartite system AB with state $\rho$.
 \end{itemize}
Apart from the von-Neumann entropy $S(\rho)$ measure of information, there also exist other measures \cite{Guo,Breuer_Laine} which depend on the particular situations at hand.\ The \textit{trace distance} measure $D(\rho^1,\rho^2)$ in Eq.\ \eqref{eq:Tr_dist} is a measure of distinguishability of two states, as discussed in Section \ref{sec:N_BLP}.\ Since more distinguishable states of a system gives more information about the states, the trace distance can thus be made an indicator of the system's information content.\ $D(\rho^1,\rho^2)$ also has the properties such as it is, like $S(\rho)$, invariant under any unitary transformation $U$ and is a contraction for any completely positive dynamical map.

 \subsection{Dynamical maps with memory}
 \subsubsection{\textit{Memoryless quantum maps}}
 If every two consecutive dynamical maps describing an open system's time evolution are independent, then the system is defined to be evolving under memoryless quantum maps.\ Thus, the Markovian dynamics, say in Eq.\ \eqref{eq:semigroup}, is an example of a memoryless quantum map or channel.\ Processes that can not be considered memoryless, are called \textit{channels with memory} \cite{Kre}.
 
 \subsubsection{\textit{Channels with memory}}
A physical map of this kind is assumed to be \textit{non-anticipatory} \cite{Man_rev}, i.e., subsequent maps do not affect previous ones.\ The part of the environment which remains coupled to the system across some consecutive time steps, is called the `\textit{memory system}' and it leads to the memory effects in the dynamics.\ Numerous examples \cite{Kre,Man_rev} of channels with memory exist.\ For example, in a \textit{localizable} \cite{Gioman} memory channel each sequential map is due to a local unitary coupling with a single multipartite correlated environment, and in a \textit{finite-memory} channel \cite{Bowman}, couplings to the memory systems last for finite times.\ 

 \subsubsection{\textit{Markovian-correlated channel}}
\label{app:Markovian}
An important example of noise with memory is the Markovian-correlated Pauli channel investigated in \cite{Macchia, Macchia_2}.\ \textit{We will consider a modification of their model in this thesis}.\ Say, we have 4 types of channels: $\phi_{i_m}$, ${i_m}={0,1,2,3}$ and they each evolve a qubit in state $\rho_{t}$ by one time step
\begin{equation}
    \rho_{t+1}=\phi_{i_t} (\rho_t) = \bigchi_{i_t}\,\rho_{t}\,\bigchi_{i_t}^{ \dagger}, \text{  with  } \bigchi_{i_t}=\sigma_{i_t}\, U
\end{equation}
where Pauli-$\sigma_i$ and $U$ produce unitary evolutions.\ The Markovian-correlated dynamical map $\Phi_T$ takes an inital state $\rho_0$ to the state
\begin{gather}
    \rho_T= \sum_{\{i_1,\ldots,i_T\}} p_{i_T|i_{T-1}}\ldots p_{i_2|i_1}p_{i_1} \  \phi_{i_T} \circ \phi_{i_{T-1}} \ldots \circ \phi_{i_1}[\rho_0]\\ 
    \text{with}\quad p_{i|j} = (1-\mu)\,p_i + \mu\, \delta_{i,j}.\
\label{eq:2} 
\end{gather}
Here $0\leq\mu\leq1$ corresponds to the \textit{relaxation time} or `\textit{memory parameter}' of the environment with $\mu=0$ for a memory-less case.
\bookmarksetup{startatroot}
\chapter{Grover's Search algorithm} 

\label{Chapter3}

\section{Quantum search algorithms}
After Grover's discovery \cite{Grover,Grover_2} of his eponymous algorithm for quantum search, several new algorithms and applications have been found based on it.\ In \cite{Vazirani} it was proved that the algorithm is optimal.\ A generalization of the algorithm, called \textit{amplitude amplification}, was devised in \cite{Brassard,amplitude}.\ There is also an important variant called `spatial search' \cite{benioff,aaronson,Shenvi} in which a marked element is searched by moving between items stored in different locations.\ In this thesis, we will focus on only the original Grover's search algorithm searching for a single element from a database.

\section{Grover's algorithm}
\label{noiseless algorithm}
\begin{figure}
    \centering
    \includegraphics[width=.75\linewidth]{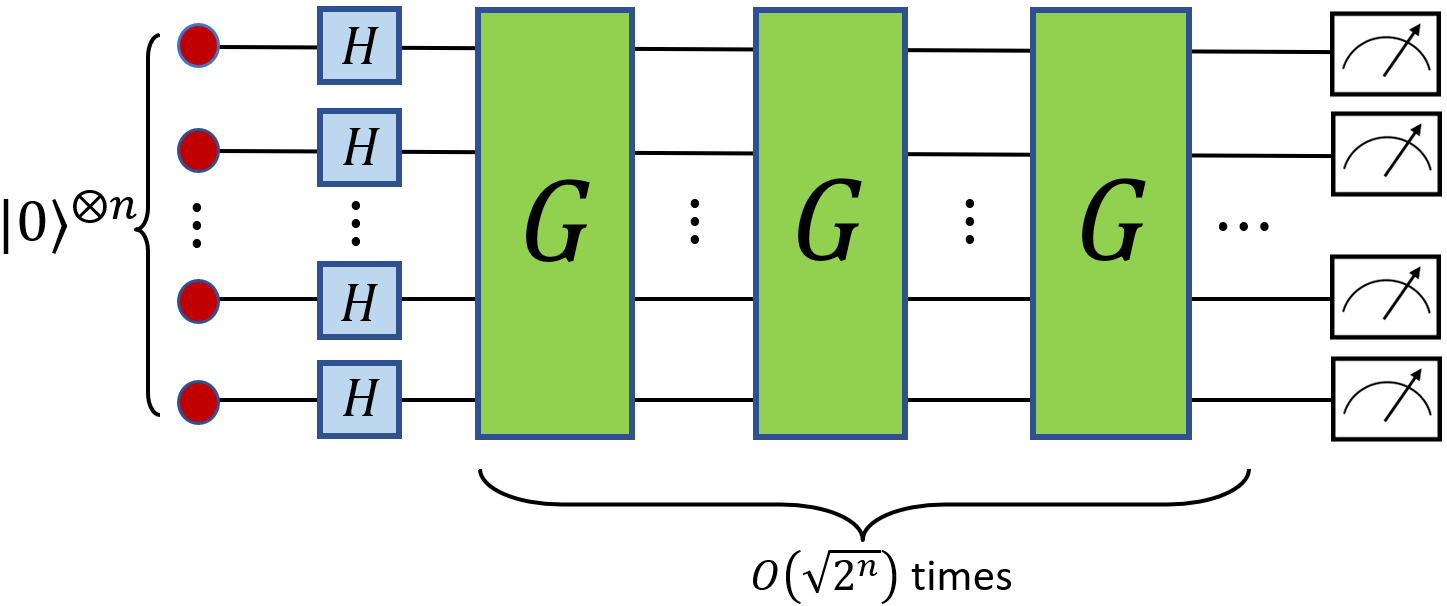}
    \caption[Quantum circuit for Grover's search algorithm]{Grover's search algorithm. The register of $n$ qubits, each $|0\rangle$, is each first subjected to a Hadamard operation.\ The second step is the operation of Grover operator, for $t=O(\sqrt{N})$ times, which is followed by a measurement on the output state.}
    \label{fig:my_label}
\end{figure}

The algorithm begins with a search space consisting of $N=2^n$ elements with $n$ being an integer.\ The elements of the space are denoted by $x=1,2,\ldots N$ and a function $f:\{0,1\}^n \rightarrow \{0,1\}$ is defined such that for the marked element $w$, $f(w)=1$ and $\forall x\neq w$, $f(x)=0$.\ To solve this problem, a classical computer evaluates $f$ for each of the elements until it gives the value 1, i.e., the marked element is found and therefore requires $O(N)$ operations.\ The advantage of Grover's search algorithm (GSA) over the classical treatment is that, by using a sequence of unitary operations, it can find the marked element by using only $O(\sqrt{N})$ queries to $f$.\ The steps of the algorithm are described as follows and a schematic demonstration is shown in Fig.~\ref{fig:my_label}. 
It starts with all qubits of an $n$-qubit register in the $|0 \rangle$ state, the eigenvector of  $\sigma_z$ operator with eigenvalue 1.\ The next step is to act on each qubit by the Hadamard operator, $H = \frac{1}{\sqrt{2}}(\sigma_x +\sigma_z)$, where $\sigma_x$ and $\sigma_z$ are Pauli operators.\ Thus the total register comes to an \textit{uniform superposition state},
 \begin{smalleralign}[\normalsize]
 |s \rangle = \left( \frac{|0\rangle +|1\rangle}{\sqrt{2}} \right)^{\otimes n} = \frac{1}{\sqrt{N}}\sum_{x = 1}^{N}|x\rangle = {\frac{1}{\sqrt{N}}}\left(\sum_{\substack{x = 1 \\ x\neq w}}^{N} |x\rangle + |w\rangle \right),
    \label{eq:unif_sup_}
\end{smalleralign}
where $|w\rangle$ is the \textit{marked state}, i.e., the state corresponding to the element we are searching for in the database of $N = 2^n$ elements.\ The state $|s\rangle$  is then acted on by the \textit{Grover operator} $G=DO$,
where $D = (2|s\rangle \langle s|- \mathds{1}_{N})$ is called the \textit{Diffuser} and $O = (\mathds{1}_{N}-2|w\rangle \langle w|)$ is the \textit{Oracle}.\ For a detailed discussion about the construction of the \textit{Diffuser} $D$, \textit{Oracle} $O$ and the Grover operator $G$, see e.g.~\cite{Kitaev,N&C}.\ The operator $G$ has the form,
\begin{equation}
    G = -\mathds{1}_N + 2|s\rangle \langle s|- \frac{4}{\sqrt{N}}|s\rangle \langle w| + 2 |w\rangle \langle w|. \label{eq:3}
\end{equation}
\begin{wrapfigure}[10]{r}{0.35\textwidth}
    \centering
    \includegraphics[width=\linewidth]{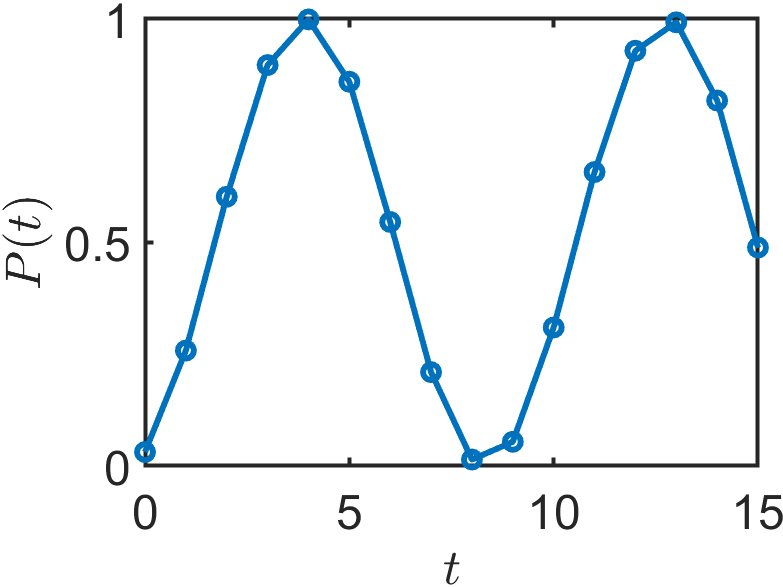}
    \caption[Success probability evolution in noiseless Grover algorithm]{Grover's algorithm for $n=5$ qubits.\ The smallest \(t\) for which \(P(t)\) is maximal is at \(t=4\).}
    \label{fig:Noiseless grover}
\end{wrapfigure}It acts on successive states until the state of the $n$-qubit register $|\psi(t)\rangle = G\,^t\, |s\rangle$
reaches close enough to the marked state $|w\rangle$.\
Here $t$ stands for the number of times the Grover operator is employed after the first step, i.e., after the Hadamard operation.\ The \textit{success probability}, i.e., the probability to find the marked state after $t^{th}$ operation, is given as $P(t) = |\langle w|\psi(t)\rangle|^2$.
It can be checked that the desired result is obtained after $t=\lfloor \frac{\pi}{4}\sqrt{N}\rfloor$ operations.\
See Fig.~\ref{fig:Noiseless grover} for the profile of the success probability with time, for a database with 32 entries.\

\vspace*{30mm}
\vspace{3cm}
\thispagestyle{empty}
\begin{center}
\begin{tikzpicture}
\node[align=center,draw,thick,minimum width=.8\textwidth,inner sep=9mm] (titlebox)%
{\fontshape{sc}\bfseries{\huge \fontsize{25}{35} Noise on Quantum Search}};
\node[fill=white] (W) at (titlebox.north) {\bfseries \HUGE \color{gray}I I};
\node (feat) at ([yshift=9mm]titlebox.north) {\textsc{\huge Part}};
\end{tikzpicture}
\vspace{.5cm}\\
\textit{This part contains the main results of the thesis.\\ In Chapter-3, the noise model is introduced and the effect of it on the discrete Grover's algorithm is discussed.\\
In Chapter-4, a physical environment is constructed that reproduces the noise in our model.\ Non-Markovianity of the process and the effect of temperature on it are then discussed.}
\end{center}
\addcontentsline{toc}{part}{Part-II: Noise on Quantum Search}
\vspace{5cm}
\epigraph{``Quantum mechanics makes no sense''}{Roger Penrose (b.\ 1931)\\\textit{Nobel Laureate in Physics}}
\bookmarksetup{startatroot}
\chapter{Searching on a noisy register} 

\label{Chapter4}
 
\textit{We consider a particular type of noise in which the ideal Grover unitaries in the noiseless GSA are modified by an additional arbitrary unitary evolution of some arbitrary qubits.\ We prove that only when these noise unitaries are the Pauli matrices, the success probability of the algorithm remains unaffected by changing the number $m\geq 1$ of noisy qubits in the register.\ We verify this result in two situations: first, a noise with no time-correlations, and second, a noise that is Markovian-correlated in time.} The results \cite{mandal2021} in this chapter can also be found in \href{https://doi.org/10.1103/PhysRevA.107.022427}{https://doi.org/10.1103/PhysRevA.107.022427}.

\section{The noise model}
\label{noise}
For a large database, 
the number of iterations of the Grover operator will be large for reaching the maximal success probability (Chapter \ref{Chapter3}) resulting in accumulation of noise or fluctuations in the circuit parameters, significantly affecting the efficiency of the algorithm.
Here we model this noise by replacing the Grover operator, \(G\), in each iteration with some probability by another operator, \(G'\), that is still unitary.\
This \emph{noisy} Grover operator is given by
\begin{equation}
\begin{split}
&G' = \bigchi_m\,G \\
   & = - \bigchi_m + 2 \left( \bigchi_m|s\rangle \right)\langle s|- \frac{4}{\sqrt{N}}( \bigchi_m|s\rangle) \langle w|\\ & \qquad \qquad \qquad \qquad \qquad \qquad \qquad + 2( \bigchi_m|w\rangle) \langle w|,
\end{split} \label{eq:6}\end{equation}
where $\bigchi_m$ is a tensor product of $n$ matrices, out of which any $m$ are each a  unitary operator, $U$, on \(\mathbb{C}^2\), and the rest $(n-m)$ are two-dimensional identity operators, $\mathds{1}_2$.\ For example, it can be
\begin{equation} \bigchi_{m}=\left(U\otimes (\mathds{1}_2)^{\otimes (n-m)} \otimes U^{\otimes (m-1)}
\right).\
\label{eq:7}
\end{equation}
Here $m$ can be considered as the ``\textit{noise strength}".\ We study how this noisy Grover operator affects the success probability and whether the success probability depends on the strength of noise.\ It is a general feature that the success probability of an algorithm reduces with increase in strength of noise, as seen e.g.
in~\cite{Biham}.\ So it will be helpful to have those noise unitaries which do not decrease the success probability with increase in the noise strength.\ We can christen such noise unitaries as ``\textit{good noise}".\
In the succeeding section, we try to identify the form of 
such good noise.\
\section{Search for ``good noise"}
\label{Proof}
To find what the good noises are, we will start with 
the most general single-qubit unitary matrix (in the computational basis),
\begin{equation}
    U = \begin{pmatrix} a & b\\ -\olsi{b} e^{i\theta} & \olsi{a} e^{i\theta}
    \end{pmatrix},
\end{equation}
with $a,b\in \mathbb{C}, |a|^2+|b|^2=1 $, $\theta \in [0,2\pi)$, and $\olsi{z}$ denoting the complex conjugate of $z$.\ The good noise corresponds to the values of $a,b$ and $\theta$, for which the success probability $P_m(t) = |\langle w|$ ${G^{\prime}}^t$ $|s\rangle|^2$ remains constant on changing the value of $m$.\ We will find the good noise by eliminating the $U$'s for which $P_m(t)$ does depend on $m$.\ We will start by finding the conditions on $U$, for keeping $P_m(t=1)$ constant with changing $m$.\ We have
 \begin{smalleralign}[\normalsize]
 P_m(1) &= \lvert\langle w|G'|s\rangle\rvert ^2 \nonumber\\
&= \left\lvert\left(1-\frac{4}{N}\right)\langle w|\bigchi_{m}|s\rangle+\frac{2}{\sqrt{N}}\langle w|\bigchi_{m}|w\rangle\right\rvert^2\nonumber\\ 
&= \frac{1}{N}\Biggl\lvert \left(1-\frac{4}{N}\right)\sum_{j=1}^{N}(\bigchi_{m})_{w,j}+2(\bigchi_{m})_{w,w}\Biggr\rvert^2 \label{eq:ch}\nonumber\\
= \frac{1}{N} &\Biggl \lvert  \left(1-\frac{4}{N}\right)(a+b)^{m-q}(\olsi{a}-\olsi{b})^q+2\, a^{m-q}\olsi{a}^q\Biggr\rvert^2,\;\;\;\;\;\;\;
\end{smalleralign}
where, $\bigchi_m |s\rangle = \frac{1}{\sqrt{N}}\begin{psmallmatrix}\sum_{j=1}^{N}(\bigchi_{m})_{1,j}\\.\\.\\\sum_{j=1}^{N}(\bigchi_{m})_{N,j}\end{psmallmatrix}$. 
It can be shown that\\
\begin{equation}
    \sum_{j=1}^{N}(\bigchi_m)_{k,j} = e^{iq\theta}(a+b)^{m-q}(\olsi{a}-\olsi{b})^q \ :=\psi_q
    \label{eq:e10}
\end{equation} \\
and $(\bigchi_m)_{k,k} = e^{iq\theta}a^{m-q}\olsi{a}^q$, 
where $q \in [0,m]$ and $q$ depends on $k$. Here each $\psi_q$ appears $\left(\frac{N}{2^m}\right)\begin{psmallmatrix}
m\\q
\end{psmallmatrix}$ times in $\bigchi_m|s\rangle$.\ 
From Eq.\ \eqref{eq:ch}, we can conclude that for getting $P_{m+1}(1) = P_m(1)$, we need either $|a|=0$ or $|b|=0$.\ Thus, for satisfying the condition $P_m(t) = P_{m+1}(t)$, we cannot have both $a$ and $b$ non-zero, and therefore we get our first condition for constructing a good noise which gives the constraints, Eqs.\ \eqref{Cond-1-1} and \eqref{Cond-1-2}.\ So, a good noise needs to obey
\begin{center}
    \textit{Condition 1:} $|a|=1$ or $|b|=1$
\end{center}

with
 \begin{numcases}{U=}
  \begin{psmallmatrix} a & 0\\ 0 & \olsi{a} e^{i\theta}\end{psmallmatrix}, \quad \text{for} \quad |a|=1,\label{Cond-1-1}\\
  \begin{psmallmatrix} 0 & b\\-\olsi{b} e^{i\theta} & 0\end{psmallmatrix}, \quad \text{for} \quad |b|=1.\label{Cond-1-2}
\end{numcases}\\
Hence $\bigchi_m$ has to be a \textit{generalized permutation matrix} for which
\begin{equation}
w^{\prime}=
\begin{cases}
 w, & \text{for} \quad |a|=1,\;\;\;\;\;\;\;\\
 \Biggl\lvert N-\frac{N}{2^m}-w+2\,\Big[w\Mod{\frac{N}{2^m}}\Big]\Biggr \rvert, & \text{for} \quad |b|=1.\;\;\;\;\;\;\;
\end{cases}
\end{equation}
 and
\begin{numcases}{\bigchi_m|s\rangle= \dfrac{1}{\sqrt{N}} \times}
    \left(\mathlarger{\sum}_{i=1}^{M}c_i \sqrt{\Delta_i}|s'_i\rangle+\alpha|w\rangle\right),&for $|a|=1$,\label{decomp_1}\;\;\;\;\\
    \left(\mathlarger{\sum}_{i=1}^{M}c_i
   \sqrt{\Delta_i}|s'_i\rangle+\alpha|w\rangle +\beta|w'\rangle\right),&for $|b|=1$,\;\;\;\;\label{decomp_2}
\end{numcases}
\begin{eqnarray*}
\text{with}\quad &&|s'_i\rangle = \frac{1}{\sqrt{\Delta_i}}\underset{\{|d_i\rangle\}\neq|w\rangle,|w'\rangle}{\mathlarger{\sum}}|d_i\rangle, \quad \Delta_i = \dim(\text{Span}\{|d_i\rangle\}),\\ 
&&\langle s'_i|s'_j\rangle = \delta_{ij}, \ \ |c_i|=1=|\alpha|=|\beta|,
\end{eqnarray*}
i.e., we get the basis $\mathbb{B}=\{|s'_1\rangle,|s'_2\rangle,\ldots,|s'_{M}\rangle,|w\rangle\}$ of dimension $(M+1)$ from Eq.~\eqref{decomp_1} and $\mathbb{B}=\{|s'_1\rangle,|s'_2\rangle,\ldots,|s'_{M}\rangle,|w\rangle, |w'\rangle \}$ of dimension $(M+2)$ from Eq.~\eqref{decomp_2} respectively.\
Our goal is to find out the noise matrices $U$ for which $P_m(t) = P_{m+1}(t),\ \forall m\geq 1$.\ 
When $m$ changes, $\bigchi_m$ changes and hence the matrix elements of $G'$ change in general.\ So, we need to find the special matrices $U$ for which the matrix elements of $G'$ doesn't change when written in the basis $\mathbb{B}$.\ There are two possibilities for a matrix $U$ of the forms in Eqs.~\eqref{Cond-1-1} and~\eqref{Cond-1-2}: the two non-zero elements are either  equal (Case $(i)$), or  unequal (Case $(ii)$).\ Case $(i)$ suggests $M=1$ and 
directly leads to the constraints, which have to be satisfied by the unitary presenting the good noise, given in Eqs.~\eqref{cond-2-1} and~\eqref{cond-2-2}.\ If we have a $U$ as in Case $(ii)$, we need to put further restrictions for the success probability to stay conserved with $m$.\ For that, since $G'$ when written in $\mathbb{B}$ must not change with $m$, we do not want $\dim(\mathbb{B})$ to change with the same.\ Thus, the number of distinct $c_i$'s in Eqs.~\eqref{decomp_1} and~\eqref{decomp_2} must remain constant with $m$.\ There are total $M$ of these coefficients.\ For $m=1$, i.e., for $\bigchi_1=U\otimes \mathds{1}_{\frac{N}{2}}$ in Case $(ii)$, there are only two distinct non-zero elements in $\bigchi_1$, because $U$ has two distinct non-zero elements.\ This implies
$M=2$.\ Since $M$ should remain constant with $m$, Case $(ii)$  leads to the restrictions for the unitaries constructing good noise to be satisfied given in Eqs.~\eqref{cond-2-3}, \eqref{cond-2-4}, \eqref{cond-2-5}. So, to summarise, we have another necessary (but not sufficient) condition:
 \begin{center}
    \textit{Condition 2:} $M$ = 1 or 2  
 \end{center}
{\small
 \begin{numcases}{\text{\normalsize Thus,}\quad \bigchi_m|s\rangle =}
 c\left(\,\sqrt{\frac{N-1}{N}}|s_1'\rangle +\frac{1}{\sqrt{N}}|w\rangle\right), & for $  |a| = 1, M=1$\label{cond-2-1}\\
 c\,\left(\sqrt{\frac{N-2}{N}}|s_1'\rangle +\frac{1}{\sqrt{N}}|w\rangle+\frac{1}{\sqrt{N}}|w'\rangle\right), & for $  |b| = 1, M=1$\label{cond-2-2}\end{numcases}}
 \begin{adjustwidth}{-30pt}{-20pt}
 {\small
 \begin{numcases}{\bigchi_m|s\rangle =}
 c_1\,\left( \sqrt{\frac{N-2}{2N}}|s_1'\rangle+\frac{1}{\sqrt{N}}|w\rangle\right)+\frac{c_2}{\sqrt{2}}|s_2'\rangle, & for $ |a|=1, M=2$ \label{cond-2-3} \\
c_1\, \left(\sqrt{\frac{N-4}{2N}}|s_1'\rangle+\frac{1}{\sqrt{N}}|w\rangle+\frac{1}{\sqrt{N}}|w'\rangle\right)+\frac{c_2}{\sqrt{2}}|s_2'\rangle, & for $ |b|=1, M=2,\alpha=\beta$ \label{cond-2-4}\\
c_1\,\left( \sqrt{\frac{N-2}{2N}}|s_1'\rangle+\frac{1}{\sqrt{N}}|w\rangle\right)+c_2\,\left(\sqrt{\frac{N-2}{2N}}|s_2'\rangle+\frac{1}{\sqrt{N}}|w'\rangle\right),& for $ |b|=1, M=2,\alpha\neq\beta$ \label{cond-2-5}\end{numcases}}
\end{adjustwidth}
So, only the $U$'s that satisfy one of the Eqs.\  \eqref{cond-2-1}-\eqref{cond-2-5}, are the unitaries corresponding to the good noise for which $P(t)$ does not depend on the number of noise sites $m$.\
 It can be shown that 
 $\psi_q$ appears $(\frac{N}{2^m})\begin{psmallmatrix}
 m\\q
 \end{psmallmatrix}$ times in the column vector $\bigchi_m|s\rangle$.\ We have the following observations.\
 \begin{enumerate}
\item[(1)] If $U$ satisfies Eq.\ \eqref{cond-2-1}, then $b=0$ and $\psi_q = c, \forall q $. Solving for $a$ and $\theta$ gives $a = e^{i\phi}= \sqrt[m]{c},\ \theta = 2\phi$, i.e., $U = \sqrt[m]{c}\begin{psmallmatrix}1 & 0\\0 & 1\end{psmallmatrix} = \sqrt[m]{c}\; \mathds{1}_2$.\ 

\item[(2)] If $U$ satisfies Eq.\ \eqref{cond-2-3}, then it turns out that we need $(a)\;\psi_q = \psi_{q+2} = c_1, \forall q$ even and $(b)\;\psi_q = \psi_{q+2} = c_2, \forall q$ odd.\ That is because, $\begin{psmallmatrix}
 m\\0
 \end{psmallmatrix}+\begin{psmallmatrix}
 m\\2
 \end{psmallmatrix}+\begin{psmallmatrix}
 m\\4
 \end{psmallmatrix}+\ldots = \begin{psmallmatrix}
 m\\1
 \end{psmallmatrix}+\begin{psmallmatrix}
 m\\3
 \end{psmallmatrix}+\begin{psmallmatrix}
 m\\5
 \end{psmallmatrix}+\ldots = 2^{m-1}$, i.e., the sum of multiplicities of elements in $\bigchi_m|s\rangle$ from the set $\{\psi_q|q\ \text{even}\}$ is equal to that in case of elements from the set $\{\psi_q|q\ \text{odd}\}$.\ Since $c_1 \neq c_2$, solving $(a)$ and $(b)$ 
for $a$ and $\theta$ give the solution, $a = \sqrt[m]{c_1},\ \theta = 2\phi-\pi$.\ The solution corresponds to $c_1=-c_2$, i.e., $U = \sqrt[m]{c}\begin{psmallmatrix}1 & 0\\0 & -1\end{psmallmatrix} = \sqrt[m]{c}\;\sigma_z$.\
\item[(3)] If $U$ satisfies Eq.\ \eqref{cond-2-2}, then $a=0$ and $\psi_q = c, \forall q $.\ Solving for $b$ and $\theta$ gives $b = e^{i\phi}= \sqrt[m]{c},\ \theta = 2\phi - \pi$, i.e., $U = \sqrt[m]{c}\begin{psmallmatrix}0 & 1\\1 & 0\end{psmallmatrix} = \sqrt[m]{c} \;\sigma_x$.\
\item[(4)] If $U$ satisfies Eqs.\ \eqref{cond-2-4} or \eqref{cond-2-5}, a similar analysis as above can be performed and the solution is $U = \sqrt[m]{c}\begin{psmallmatrix}0 & 1\\-1 & 0\end{psmallmatrix} = \sqrt[m]{c}\;i\;\sigma_y$.\

 \end{enumerate}
Here $\sqrt[m]{c}$ is only a constant phase factor.\ We can see from the above discussion that the restricted set of unitary qubit evolutions that are candidates for being good noise, are the matrices $e^{i\phi}\,\mathds{1}_2$, $e^{i\phi}\,\sigma_x$, $e^{i\phi}\,\sigma_y$ and $e^{i\phi}\,\sigma_z$, for any $\phi \in [0,2\pi)$ for $t=1$.\
We will now check if this set of noise unitaries are `good noise' for all times $t>1$.\ Similar to Eq.\ \eqref{eq:ch} we can write for $t=2$,
 \begin{smalleralign}[\normalsize]
P_m(2) &= \lvert\langle w|G'\,^2|s\rangle\rvert ^2 \nonumber\\
 &= \Bigg\lvert\left(1-\frac{4}{N}\right)\left[\left(2 \langle s| \bigchi_m |s\rangle  - \frac{4}{\sqrt{N}}\langle w|\bigchi_m|s\rangle + 2\langle w|\bigchi_m |w\rangle \right)\langle w|\bigchi_m|s\rangle-\frac{1}{\sqrt{N}}\right] +\nonumber\ \ \ \  \ \ \ \\& \quad \quad \quad \quad +\frac{2}{\sqrt{N}}\left[2\langle w|\bigchi_m|s\rangle \langle s|\bigchi_m|w\rangle -1 - \frac{4}{\sqrt{N}}\langle w|\bigchi_m |s\rangle \langle w|\bigchi_m|w\rangle+2\langle w|\bigchi_m|w\rangle^2\right]   \Bigg\rvert^2, \nonumber \label{Pm2}\\ 
\end{smalleralign}
where $\bigchi_m ^2 = \mathds{1}$ for a Pauli matrix $U$.\ 
It can be shown that
\begin{equation}
   \langle s|\bigchi_m|s\rangle =\frac{1}{N}\sum_{k=1}^{N}\psi_q = \frac{2^{(n-m)}}{N} [(a+b)+e^{i\theta}(\olsi{a}-\olsi{b})]^m.\  
\end{equation}

Now, for $U=\sigma_x$, we have $a=0$, $b=1$, and $\theta=\pi$.\ So, $\langle s|\bigchi_m|s\rangle = 1$, $\langle w|\bigchi_m|s\rangle =\frac{1}{\sqrt{N}}$ and $\langle w|\bigchi_m|w\rangle = 0$, $\forall m$.\ Putting these values in Eq.\ \eqref{Pm2}, we can see that $P_m(2)$ is independent of $m$.\ 

For $U=\sigma_z$, we have $b=0$, $a=1$ and $\theta=\pi$. So, $\langle s|\bigchi_m|s\rangle = 0$, $(\langle w|\bigchi_m|s\rangle)^2$, $\langle w|\bigchi_m |w\rangle \langle w|\bigchi_m|s\rangle$ 
remains constant, $\forall m$.\ Therefore from Eq.\ \eqref{Pm2}, our claim for $\sigma_z$ to be a good noise also holds for $t=2$.\

In case of $U=\sigma_y$, $a=0$, $b=-i$ and $\theta=\pi$.\ So we have $\langle s|\bigchi_m|s\rangle=0$, $\langle w|\bigchi_m|w\rangle=0$ and $\langle w|\bigchi_m|s\rangle
= \frac{1}{\sqrt{N}} e^{iq\pi}\, (-i)^{m}$.\ Hence,
 \begin{smalleralign}[\normalsize]
 P_m(2)=&\ \frac{1}{N}\Big\lvert\left(1-\frac{4}{N}\right)\left[ \left(2\, \langle w|\bigchi_m|s\rangle\right)^2 +1\right]\nonumber\\ &\quad \quad \quad \quad \quad -4\lvert\langle w|\bigchi_m|s\rangle\rvert^2 +2 \Big\rvert^2\nonumber\\
    &= \frac{1}{N}\Big\lvert\frac{1}{N}\left(1-\frac{4}{N}\right)\left(2\, e^{iq\pi}\, (-i)^{m}\right)^2  - \frac{8}{N} +3 \Big\rvert^2\nonumber\\
    &= \frac{1}{N}\Big\lvert\frac{4}{N}\left(1-\frac{4}{N}\right)(-1)^{m}  - \frac{8}{N} +3 \Big\rvert^2.
    \label{eq:e24}
\end{smalleralign}
From Eq.\ \eqref{eq:e24} we can infer that for $U=\sigma_y$, the success probabilities $P_m(2)=P_{m+2}(2), \ \forall m$, i.e., although for any consecutive $m$, the success probability is not constant, it does remain the same for $m$ staying either odd or even.\ We have not analytically shown here if this is true for $t>2$ in case of $\sigma_y$.\ But in Fig.\ \ref{fig:s_Y}, it is shown that indeed $P_m(t)=P_{m+2}(t), \ t\geq 1, \ \forall m$.\

For any $t\geq 2$, $P_m(t)=|\langle w|G'\,^t|s\rangle|^2$ consists of the terms $\langle s|\bigchi_m|s\rangle$,\ $\langle w|\bigchi_m|s\rangle$,\ $\langle w|\bigchi_m|w\rangle$ or combinations of these terms.\ Since it turns out that for $\sigma_x$ and $\sigma_z$, $P_m(t)$ are independent of $m$, we have $P_m(t)=P_{m+1}(t), \forall m$.\

So for example, if we have total of $N=8$  elements in a search space, then it turns out that the evaluation of the success probability in case of $m = 2$ noise sites and that in case of $m=3$ noise sites will be indistinguishable if the qubits in those sites are rotated by the good noises, i.e., $U \in \{\sigma_x,\sigma_z\}$.\ The success probabilities in the cases where $m=2$, $m=4$ or $m=6$, will be exactly the same in case of $U=\sigma_y$.\ Similarly, the cases of $m=1$, $m=3$ or $m=5$ will be indistinguishable when $U=\sigma_y$.\ It is to be noted that there may be some unitary $U$, other than these Pauli matrices, which makes the success probability independent of $m$ for some particular time $t$ and not at other times.\ The Pauli matrices $\sigma_x$ and $\sigma_z$ are special in the sense that when $U$ is one of these, the success probability becomes independent of $m$, \textit{for all \(t\)}.\

Another important observation is that none of the conditions used above put restrictions on what the positions of the $m$ unitaries are, out of the total $n$ positions.\ The coefficients $c_i$ remain the same for any arrangement of the $m$ unitaries.\ So, as mentioned earlier in Sec.~\ref{noise}, the success probability does not depend on the positions of the qubits which evolve under the noise unitary $U\in \{\sigma_x, \sigma_y, \sigma_z\}$.\ This result is also depicted in Fig.\ \ref{fig:Comparison_sites}.

\section{Markovian-correlated noise}
\label{Example}
\begin{figure}[h]
    \centering
    \includegraphics[width=.7\linewidth]{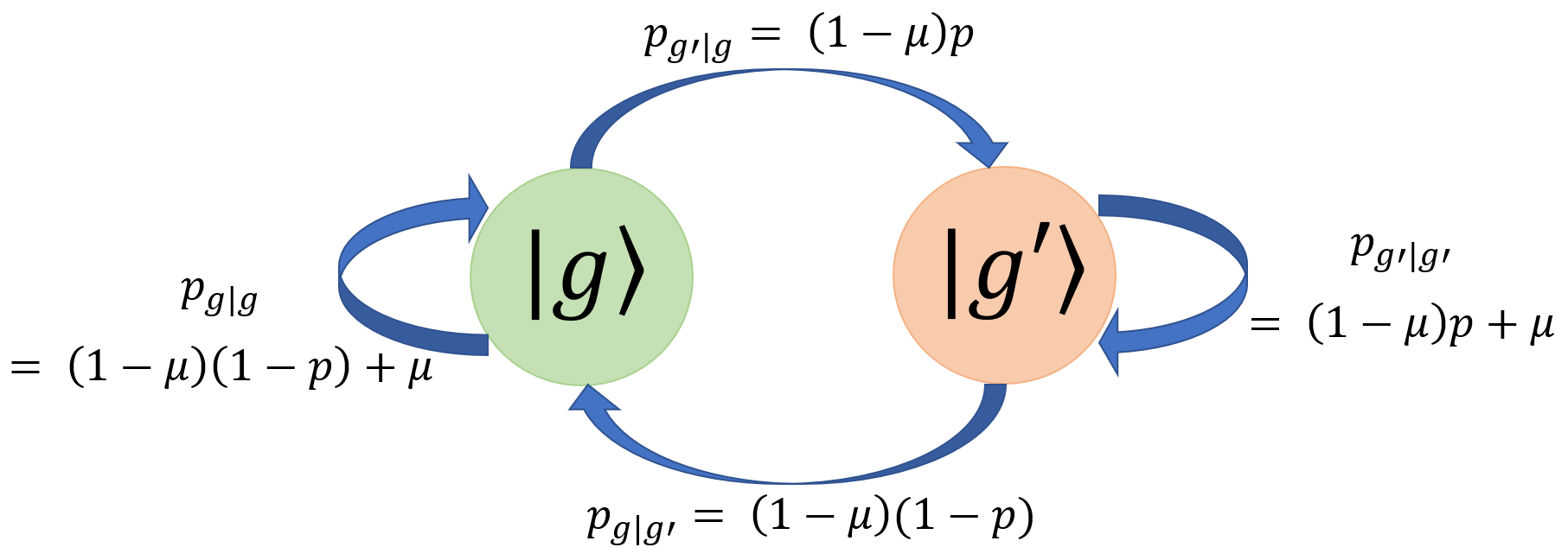}
    \caption[Schematic diagram of Markovian transition probabilities]{Schematic diagram of Markovian transition probabilities for $t \ge 2$.\ The four probabilities corresponding to  application of the ideal and noisy Grover operators are given by $p_{g|g}$, $p_{g^{\prime}|g}$, $p_{g|g^{\prime}}$ and $p_{g^{\prime}|g^{\prime}}$.\ Please see the text for details.}
    \label{fig:Markovian probabilities}
\end{figure}

To illustrate the above results, we will consider the situation where the noise is Markovian-correlated in time (see the discussion in Section \ref{app:Markovian}).\ This potentially important variety of noise with memory 
has not yet been studied before in case of GSA.\ It is also to be noted here that our results are not exclusive to only this kind of noise.\ In~\cite{Biham}, they examined the effect of noise on GSA incorporating the noise in the Hadamard gate in the first step of the search algorithm.\ The noise was uncorrelated, i.e., the noisy unitary Hadamard gates were constructed in a completely arbitrary manner from a Gaussian distribution and the unitary at each step did not depend on the unitary in the previous step(s).\ In contrast, we consider a correlated noisy Grover operator which at each iteration probabilistically depends on the preceding one.\ 
It is easy to understand the situation by considering the situation where any $m$ of the $n$ qubits become connected to another degree of freedom which we call as the \textit{walker}.\ A schematic diagram is shown in Fig.~\ref{fig:Markovian probabilities}.\ The walker has two orthogonal states $|g\rangle$ and $|g'\rangle$, and when it is in $|g'\rangle$, all the $m$ qubits connected to it are rotated by a unitary $U$ and the other $(n-m)$ are left as they were.\ Thus, with each time step (iteration), application of an ideal unitary Grover operator $G$ is followed by one of the following:
\begin{enumerate}
    \item[($I$)] any $m$ out of $n$ qubits are rotated by a unitary $U$, i.e., walker is in state $|g'\rangle$, or,
    \item[($II$)] all the $n$ qubits are left untouched, i.e., walker is in state $|g\rangle$.\ 
\end{enumerate}

The transition probabilities for $|g\rangle \leftrightarrow |g'\rangle$ are determined by a dichotomous Markov chain considered in~\cite{Macchia}, and described by Eq.~\eqref{eq:2} (in Section~\ref{app:Markovian}).\ To make the situation clearer, let us assume after the $(t-1)^{\text{th}}$ Grover iteration, the $n$ qubit register is in a state given by the density matrix $\rho_{t-1}$.\ Thus, \textit{before} the \((t+1)^{\text{th}}\) iteration, the register can be in the following two possible states:
\begin{eqnarray*}
&(I)& \ \rho_t = G'\,\rho_{t-1}\,G'^{\dagger},\;\; \text{with}\;\; G'= \bigchi_{m}\,G,\\
&(II)& \ \rho_t = G\,\rho_{t-1}\,G^{\dagger}.
\end{eqnarray*}
At $t=1$, i.e., on
the first Grover iteration, the probabilities of $(I)$ and $(II)$ are determined by the initial probabilities of the walker to be in states $|g'\rangle$ and $|g\rangle$ respectively.\ These probabilities are called \textit{stationary probabilities} and are taken to be $p_{g'} = p$ and $p_g = (1-p)$ respectively.\ Here $p$ can be referred to as the \textit{noise probability}.\ 
At any later time $t$, the probabilities are determined by the $(t-1)^{th}$ iteration and the \textit{memory} parameter, $\mu$.\ That is, for $t \geqslant 2$,
    \begin{numcases}{p_{k|l}=}
    (1-\mu)\,p_l + \mu, & \text{for} \quad $k=l$ \label{eq:8}\\
    (1-\mu)\,p_{k}, & \text{for} \quad $k \neq l$.\label{eq:8_1}
    \end{numcases}
where $k$, $l$ can be $G$ or $G^{\prime}$.\ See Section~\ref{app:Markovian} for further discussions.\

Before the application of the first Grover iteration, the $n$-qubit register is in the uniform superposition state $|s\rangle$.\ Let us denote the density matrix corresponding to this state as $\rho_0 \coloneqq |s\rangle  \langle s| $.\
So, the density matrix of the composite system containing the walker and the register before applying any Grover iteration is given as $R_0 = \left(\frac{|g\rangle+| g'\rangle}{\sqrt{2}}\right)\left(\frac{\langle g|+\langle g'|}{\sqrt{2}}\right) \otimes |s\rangle\langle s|$.\ So, the state of the register $\rho_1$ after the first Grover iteration will be $\rho_1 =$ $\text{Tr}_{walker}$ $\{R_1\}$ $=p_{g}\Phi^0[\rho_0]+p_{g'}\Phi^1[\rho_0]$ and the state after the second Grover iteration, $\rho_2 =$ $\text{Tr}_{walker} \{R_2\}$ $=p_{g}p_{g|g}\ \Phi^0[\Phi^0[\rho_0]]$ $+ p_{g}p_{g|g'}\ \Phi^0[\Phi^1[\rho_0]]$ $
    + p_{g}p_{g'|g}\ \Phi^1[\Phi^0[\rho_0]]$  $+$  
    $p_{g'}p_{g'|g'}$ $\Phi^1[\Phi^1[\rho_0]]$, where $R_1=S_0R_0$ and $R_2=S\,R_1$, with $S_0$ and $S$ being the transition operators given by
\begin{smalleralign}[\normalsize]
S_0 &=\Biggl(p_g|g\rangle\langle g| \otimes \Phi^0[.] + p_{g'}|g'\rangle\langle g'| \otimes \Phi^1[.]
+p_g|g\rangle\langle g'| \otimes \Phi^0[.] + p_{g'}|g'\rangle\langle g| \otimes \Phi^1[.]\Biggr),\\
S &=\Biggl(p_{g|g}|g\rangle\langle g| \otimes \Phi^0[.]+p_{g|g'}|g\rangle\langle g'| \otimes \Phi^0[.] +p_{g'|g} |g'\rangle\langle g| \otimes \Phi^1[.] + p_{g'|g'}|g'\rangle\langle g'| \otimes \Phi^1[.]\Biggr),
\end{smalleralign}
where $\Phi^0[\rho] = G \rho \, G^\dagger$ and $\Phi^1[\rho] = G' \rho \, G'^\dagger$. Similarly, for $t\geqslant2$, we have
\begin{equation}
    R_t = S\ ^{t-1} R_1,\quad   \rho_t = \text{Tr}_{walker}\{R_t\}.
\end{equation}
The success probability, i.e., the probability to find the marked state at time $t$, is given as
\begin{equation}
    P(t) = |\langle w|\rho_t |w\rangle|.
    \label{29}
\end{equation}

This kind of correlated noise with partial memory can potentially be found in real quantum devices, and it has been shown to provide an enhancement in the transmission of classical information as compared to transmission through noisy channels without memory~\cite{Macchia}.\ We investigate the effects of this Markovian correlated noise on the GSA numerically, and the results are gathered in the  subsection below.\

\subsection{A special case}
\label{App_A}
In case of $U = \sigma_x$ from Eq.\ \eqref{eq:6}, we get  $\bigchi_m |s\rangle = |s\rangle$.\ So, we can express all the states in terms of the orthogonal basis vector set $\{|\bar{s}\rangle, |w\rangle, |w'\rangle\}$, with
\begin{equation*}
|\bar{s}\rangle \coloneqq \frac{1}{\sqrt{N-2}}\sum_{\substack{x = 1 \\ x\neq w,w'}}^{N}|x\rangle.\
\end{equation*}
The uniform superposition state $|s\rangle$ then becomes
$\begin{psmallmatrix}
  \sqrt{\frac{N-2}{N}}, & \frac{1}{\sqrt{N}}, & \frac{1}{\sqrt{N}}
\end{psmallmatrix}^{\dagger} \label{3}
$ in this basis.\ Hence $G$ and $G^\prime$ have the forms,
\begin{equation} 
G=\Biggr(\begin{smallmatrix} 
2\frac{N-2}{N}-1 & -2\frac{\sqrt{N-2}}{N} & 2\frac{\sqrt{N-2}}{N}\\
2\frac{\sqrt{N-2}}{N} & -\frac{2}{N}+1 & \frac{2}{N}\\
2\frac{\sqrt{N-2}}{N} & -\frac{2}{N} & \frac{2}{N}-1
\end{smallmatrix}\Biggr), \quad G' = \Biggr(\begin{smallmatrix} 
2\frac{N-2}{N}-1 & -2\frac{\sqrt{N-2}}{N} & 2\frac{\sqrt{N-2}}{N}\\
2\frac{\sqrt{N-2}}{N} & -\frac{2}{N} & \frac{2}{N}-1\\
2\frac{\sqrt{N-2}}{N} & -\frac{2}{N}+1 & \frac{2}{N}
\end{smallmatrix}\Biggr)
\label{eq-4.15}
\end{equation}
It is evident from the expressions above that, at least for the case $U = \sigma_x$, although changing $m$ does change the forms of the basis vectors $|w'\rangle$ and $|\bar{s}\rangle$ in the computational basis, elements of all the states or operators like $|s\rangle$ or $G'$ remain the same in 
$\{|\bar{s}\rangle, |w\rangle, |w'\rangle\}$ basis.\ Thus, increasing or decreasing the noise strength $m$ does not affect the success probability of the algorithm in case of $U=\sigma_x$ and $m\geq 1$.\

\subsubsection{Success probability for perfect memory and \texorpdfstring{$U=$}{} \texorpdfstring{$\sigma_x$}{}}
\label{perfect memory}
Here, we consider the case when $\mu = 1$, i.e., the \textit{perfect memory}. On the first iteration (i.e., $t=1$), $G$ occurs with probability $(1-p)$ and $G'$ with $p$.\ Let us assume at $t = 1$, $G$ is applied.\ Due to perfect memory, for all $t\geqslant 2$, the same operator $G$ will be applied.\ This scenario corresponds to an ideal \textit{noiseless} GSA and marked state is reached at $t \approx \frac{\pi}{4}\sqrt{N}$ \cite{N&C}.\ Instead if $G'$ is applied at $t=1$, for $t\geqslant2$ the state of the whole $n$-qubit register would be $|\psi(t)\rangle = G'\, ^t\ |s\rangle$.\ To find the explicit expression for $G'\,{}^t$, we diagonalize $G'$ to get $G'_d$ and $G'_d {}^t$.\ Thus, using the explicit form in Eq.\ \eqref{eq-4.15} for $U = \sigma_x$,
 \begin{smalleralign}[\normalsize]
    \langle w|G'\,^t|s\rangle = \langle w|XG'_d\,^t X^{-1}|s\rangle 
    =(-1)^{t+1} \frac{1}{\sqrt{N}}\, \cot\left(\frac{\theta}{2}\right)
    \times \operatorname{\mathbb{I}m}\left[\tan\left(\frac{\theta}{2}\right)\left(\tan\left(\frac{\theta}{2}\right)-i\ \right)e^{i t \theta}\right]
\end{smalleralign}
where $X$ is the diagonalizing matrix, $\theta = \cos^{-1}(\frac{2}{N})$ and $\operatorname{\mathbb{I}m}[\cdot]$ denotes the imaginary part of a complex number.\ Then the probability to find the marked state at time $t$, is given as
\begin{equation}
    P(t) = |\langle w|G'\,^t|s\rangle|^2
     = \frac{1}{N} \cos^2 (\theta t)\left(\tan\left(\frac{\theta}{2}\right)\tan(\theta t)-1\right)^2.\ \label{C6}
\end{equation}
The first maximum of $P(t)$ in Eq.\ \eqref{C6} is analytically found at $t = \left(\frac{\pi}{\theta}-\frac{1}{2}\right) \approx \left(\frac{3}{2}+\frac{8}{\pi N}\right)$.\

\subsection{Patterns of the success probability}
\label{numerics}

In this subsection, we will see that the invariance of the success probabilities in case of Pauli noise unitaries, persists irrespective of any time-correlation in the noise.\
\subsubsection{\textit{Noise without memory}}
\begin{figure}[h!]
    \includegraphics[width=0.7\linewidth]{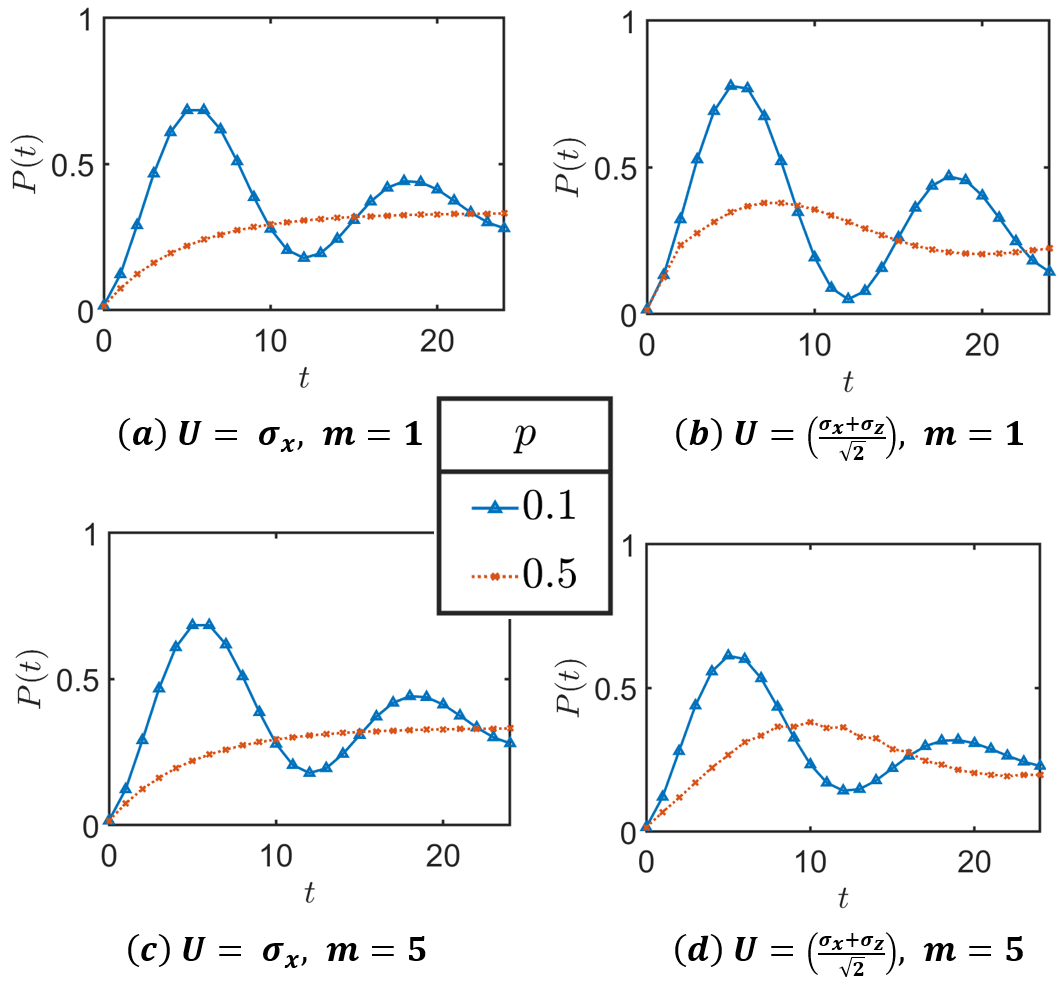}
        \centering
    \caption[Comparison of success probabilities for different noise unitaries $U$ in case of $\mu=0$, $n=6$]{Success probabilities of GSA
    in presence of noise without any time-correlation. So, $\mu=0$ here.\ We took \(n=6\).\ $P(t)$ is on the vertical axis and time along the horizontal axis.\ The inset shows values of the noise probability $p$.\ The plots are for different \(U\) and \(m\) as displayed below each plot.}
    \label{fig:Memoryless}
\end{figure}
\begin{figure}[h!]
    \includegraphics[width=0.6\linewidth]{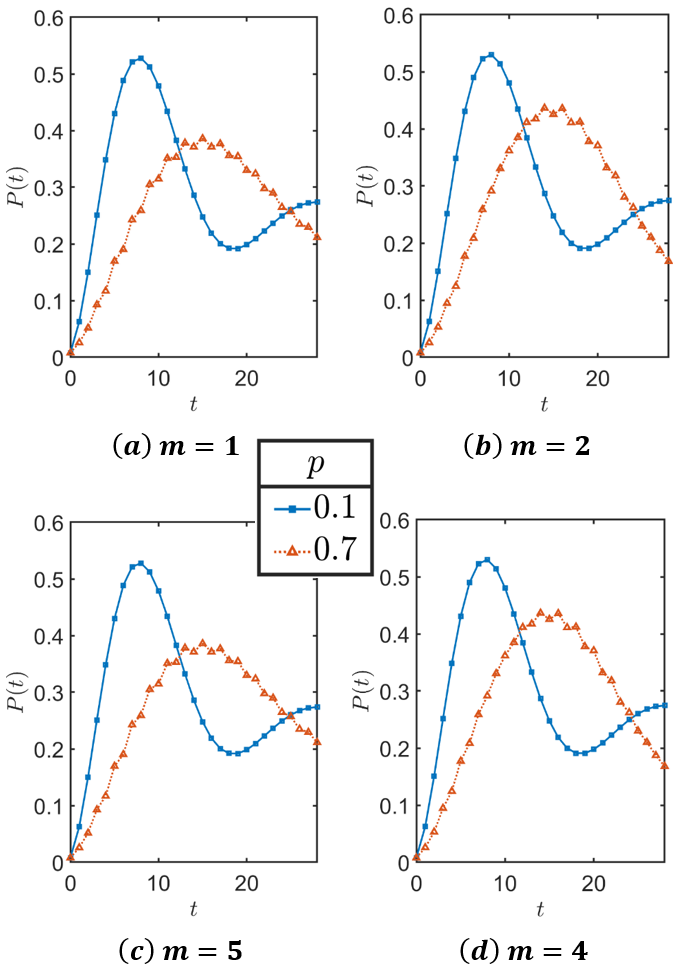}
    \centering
    \caption[Success probabilities for $\mu=0$, $n=7$, $U=\sigma_y$]{Success probabilities of GSA
    in presence of noise without any time-correlation, i.e., $\mu=0$.\ The plots are for \(n=7\) and $U=\sigma_y$.\
    The inset table exhibits the symbols used in the plots for noise probability $p=0.1$ and $p=0.7$.
    The different plots are for different noise strengths \(m\), as displayed below each plot.}
    \label{fig:s_Y}
\end{figure}

The case of $\mu=0$ in Eqs.\ \eqref{eq:8}, \eqref{eq:8_1}, leads to noise without any memory or time-correlation.\ So at each time step, the probability for the Grover operation to become noisy is $p_{g'}=p$.\ In Fig.\ \ref{fig:Memoryless}, we see that the success probability's evolution, $P(t)$, for a given noise probability is unchanged when the number of noisy qubits is increased from $m=1$ to $m=5$ for $U=\sigma_x$.\ We contrast this with the change in $P(t)$ in case of $U=(\sigma_x+\sigma_z)/\sqrt{2}$, i.e., the Hadamard operator.\ 
We have plotted the success probability's evolutions for $m=1,\ 2,\ 4,\ 5$ in case of the noise unitary $\sigma_y$ and $\mu=0$ in Fig.\ \ref{fig:s_Y}.\ As discussed in Sec.\ \ref{Proof}, the behaviour is exactly the same for odd noise strengths, i.e., for $m=1$ and $m=5$.\ The same is true for even noise strengths $m = 2$ and $m=4$.

\subsubsection{\textit{Noise with finite time-correlation}} 
\begin{figure}[t]
    \includegraphics[width=.62\linewidth]{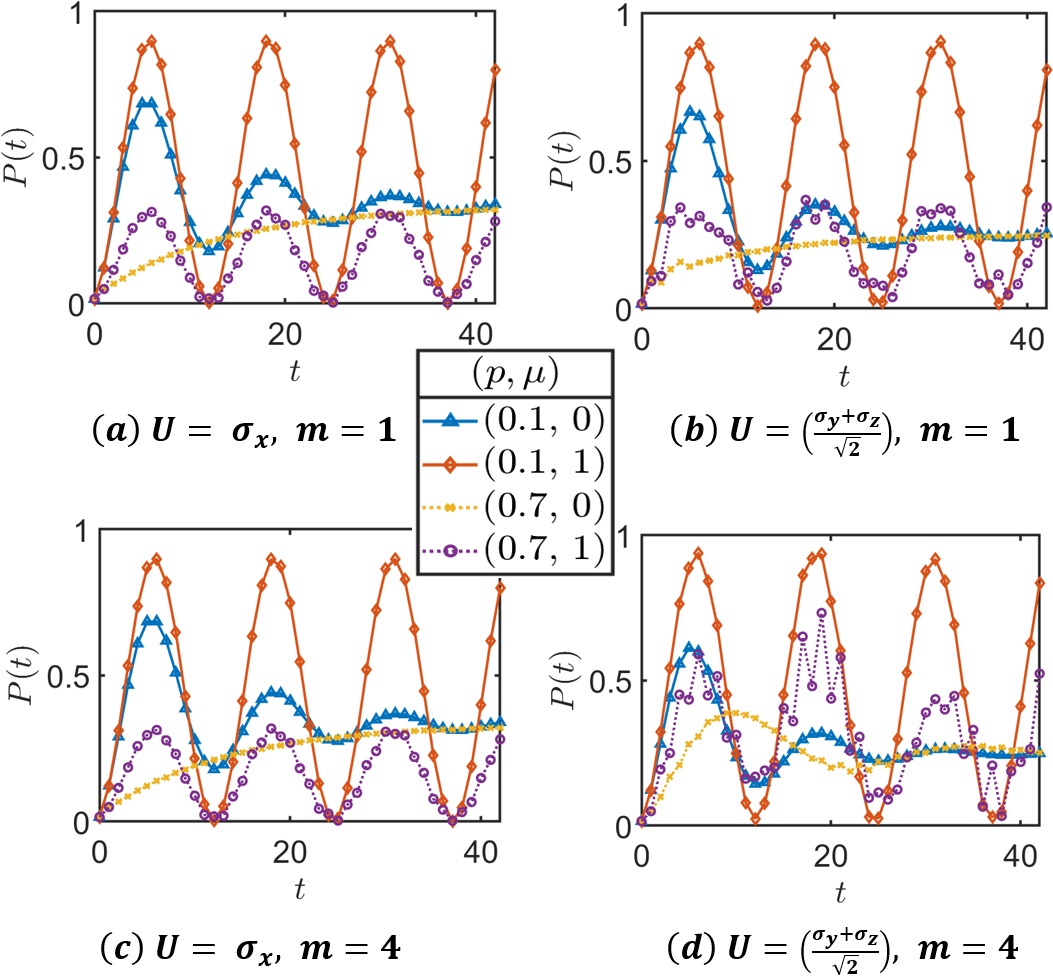}
        \centering
    \caption[Success probabilities of GSA
    in presence of noise]{Success probabilities of GSA
    in presence of noise.\ The plots are for \(n=6\).\ Here we have plotted $P(t)$ on the vertical axis and the number of Grover iterations along the horizontal axis.\
    The inset table exhibits the symbols used in the plots for different pairs of values of the noise probability $p$ and memory parameter $\mu$.\
    The different plots are for different \(U\) and \(m\), as displayed below each plot.}
    \label{fig:Success probabilities}
\end{figure}

The success probability $P(t)$ of GSA for non-zero (positive) memory $\mu$ and for $n=6$ qubits for two different noise unitaries are depicted in Fig.~\ref{fig:Success probabilities}.\
  Here we have used the form of noise as $\bigchi_m= U^{\otimes m}\otimes \mathds{1}_2 ^{\otimes (6-m)}$ with $m=1$ and $4$.\ We can observe that the success probability $P(t)$ depends on the noise probability $p$ and the memory parameter $\mu$.\ It is obvious that the success probability reduces with increasing noise probability, and we can see from all the four panels that for a very high noise probability, the oscillatory behaviour of $P(t)$ tends to vanish.\ An interesting observation from a comparison of panels $(a)$ and $(c)$ is that they support our results regarding the algorithm's behaviour under a ``good'' noise.\ It can be seen that for the Pauli matrix $U=\sigma_x$, $P(t)$ for a given $p$ and $\mu$, remains unaffected when we change the number of noise sites $m$ on which $U$ is applied.\ Whereas for a non-Pauli unitary $U=(\sigma_y+\sigma_z)/\sqrt{2}$, which was shown not to be a good noise before, the success probability $P(t)$ changes with the noise strength $m$.\ Compare panels $(b)$ and $(d)$.\ We can see that for memory-less ($\mu=0$) noise at a very high noise probability say, $p=0.7$, the search algorithm becomes completely inefficient as the original oscillating nature of success probability completely disappears.\ The GSA is thus more efficient when the noise has higher memory and thus the correlation of the noisy unitaries are found to be more beneficial than the noise without memory.
\begin{figure}[h!]
    \centering
    \includegraphics[width=.6\linewidth]{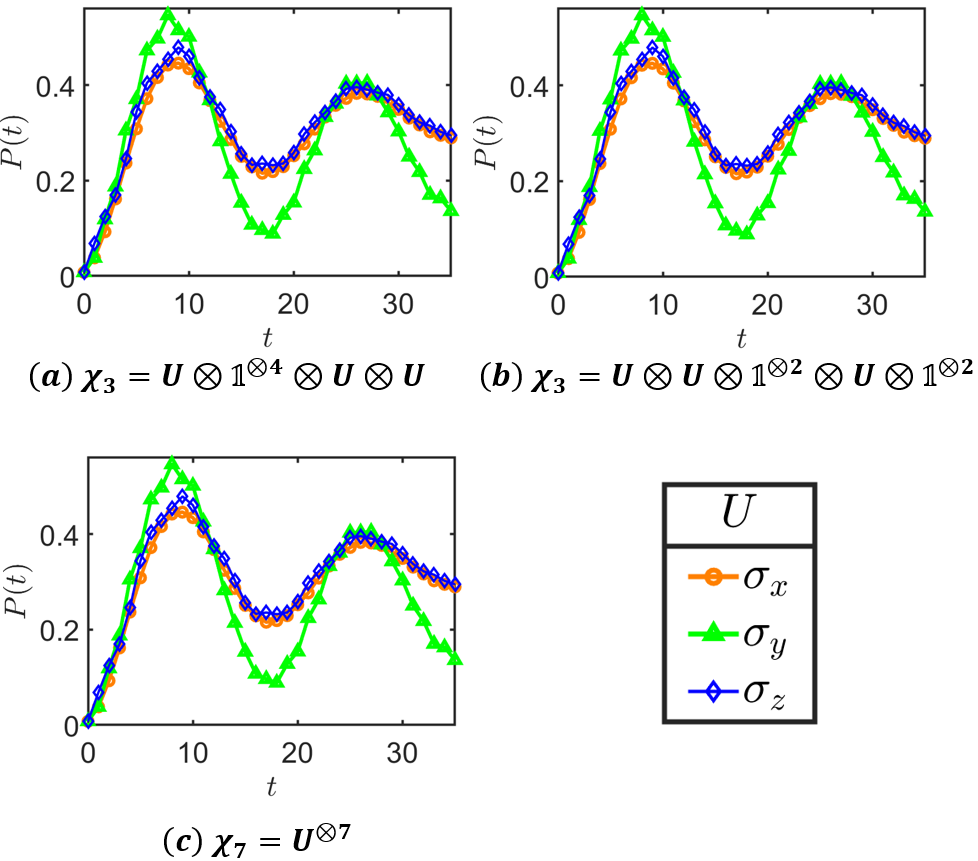}
    \centering
    \caption[Success probabilities of GSA for good noises]{Case of Good noises.\ Here \(n=7\), $p=0.5$ and $\mu=0.9$. The $\chi$'s are displayed below each plot.\ The corresponding $U$'s are in the inset.}
    \label{fig:Comparison_sites}
\end{figure}

Moreover for higher values of $p$ for which the oscillation of $P(t)$ completely vanishes, the correlated noise plays an advantageous role.\ If we compare the lines corresponding to $(p,\mu)=(0.7,0) \ \text{and} \ (0.7,1)$ in the figure, we can find that the oscillatory nature reappears for a higher memory parameter and a better success probability can be obtained for a smaller number of iterations of the operator.\ \begin{figure}[t]
    \centering
    \includegraphics[width=.7\linewidth]{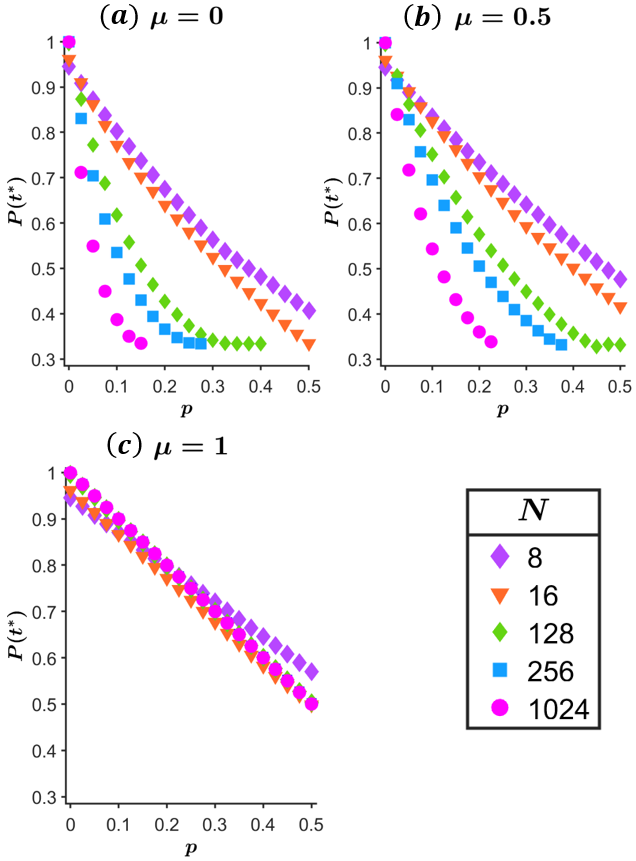}
    \caption[Effects of memory, database size and noise probability on the algorithm's success probability]{Effects of memory, database size and noise probability on the algorithm's success probability.\ We have presented here the values of $P(t^*)$, with respect to the noise probability $p$, for different $\mu$'s.\ The algorithm is performed on $n=\log_2 N$ qubits.\ Here the noise unitary considered is $U=\sigma_x$.\ All quantities used are dimensionless.\ }
    \label{fig:P_first max}
\end{figure}
The analysis for the evolution of success probability under $U=\sigma_x$ in case of perfect memory is summarised in Section \ref{perfect memory}.\
The success probabilities for the noiseless situation ($U=\mathds{1}_2$) and the good noises from Sec.~\ref{Proof} are plotted with respect to the number of iterations of the noisy Grover operator in Fig.~\ref{fig:Comparison_sites}.\
As we have commented previously, the positions of the noisy qubits do not matter if $U$ is one of the Pauli matrices.\
It is clear from the figure that for a given $U$, $P(t)$ remains unchanged while changing the positions of the noise sites.\
Fig.~\ref{fig:P_first max} gives an overview of the effects of memory, database size and noise probability on the algorithm's success.\ Here we plot the success probabilities at their first maxima $P(t=t^*)$ with respect to the noise probability $p$ for $U=\sigma_x$.\ The effect of memory is contrasted in the three subplots.\ As observed in Fig.~\ref{fig:Success probabilities}, here also we can see that for a given value of $p$, a higher memory $\mu $ of the noise provides a higher success probability of the algorithm.\

\bookmarksetup{startatroot}
\chapter{Non-Markovianity \& Thermal effects} 
\textit{In this chapter we introduce a collisional model \cite{ciccarello,Esposito,Campbell,Salvatore,Bodor,Man} that exactly reproduces the time evolution of our system under a noise with memory (analyzed in Chapter-3).\ The unitary governing the collisions is decomposed to point out the origin of the conditional probabilities that characterize our Markovian-correlated noise.\ We define a measure of CP-divisibility for the discrete-time evolution and show that although the system's evolution is non-Markovian for most of the parameter space, the back-flow of information from the environment happens only within a subspace of all $\{p,\mu\}$ values.\ By introducing an elementary effect of temperature in the model, we find that increasing temperature leads to increasing information drainage, i.e., decreasing non-Markovianity of the process.  }
\label{Chapter5}
\section{Noisy evolution as a sequence of collisions}
\subsection{Kraus representation}
It turns out that the evolution of the noisy register as described in Section-\ref{Example} has an alternate representation.\ Starting with the same initial (walker+system) state $R_0 = \left(\frac{|g\rangle+| g'\rangle}{\sqrt{2}}\right)\left(\frac{\langle g|+\langle g'|}{\sqrt{2}}\right) \otimes |s\rangle\langle s|$, the following Kraus decomposition can be given for the evolution from $R_0$ to $R_1$
\begin{gather}
    R_1 = \sum_{\alpha=1}^{4}\ K_{\alpha}\ R_0\  K_{\alpha}^{\dagger}\label{eq:K_R1}\\ 
    K_1 =  \sqrt{p_g} \begin{psmallmatrix} G & 0\\ 0 & 0\end{psmallmatrix}, K_2 =  \sqrt{p_g} \begin{psmallmatrix} 0 & G\\ 0 & 0\end{psmallmatrix}, K_3 =  \sqrt{p_{g'}} \begin{psmallmatrix} 0 & 0\\ G' & 0\end{psmallmatrix}, K_4 =  \sqrt{p_{g'}} \begin{psmallmatrix} 0 & 0\\ 0 & G'\end{psmallmatrix}.
\end{gather}
Thus the evolution from $t=0$ to $t=1$ is given by a CP map that is also TP since $\sum_{\alpha}\ K_{\alpha}^{\dagger}\ K_{\alpha} = \mathds{1}$, but \textit{non-unital}, $\sum_{\alpha}\ K_{\alpha}\  K_{\alpha}^{\dagger} \neq \mathds{1}$. This process takes the system state to $\rho_1 = $ $\text{Tr}_{walker}\{R_1\}$ $=p_{g'}\Phi^0[\rho_0]+p_{g}\Phi^1[\rho_0]$ which is the same as we obtained before.\\
The (walker+system) states for $t\geq 2$ can be obtained likewise
\begin{gather}
    R_t = \sum_{\alpha=1}^{4}\ K_{\alpha}\ R_{t-1}\  K_{\alpha}^{\dagger} \label{eq:K_Rt}\\
    K_1 =  \sqrt{p_{g|g}} \begin{psmallmatrix} G & 0\\ 0 & 0\end{psmallmatrix}, K_2 =  \sqrt{p_{g|g'}} \begin{psmallmatrix} 0 & G\\ 0 & 0\end{psmallmatrix}, K_3 =  \sqrt{p_{g'|g}} \begin{psmallmatrix} 0 & 0\\ G' & 0\end{psmallmatrix}, K_4 =  \sqrt{p_{g'|g'}} \begin{psmallmatrix} 0 & 0\\ 0 & G'\end{psmallmatrix}.
\end{gather}
where the $p_{k|l}$ are the conditional probabilities (see Section~\ref{app:Markovian})
 \begin{numcases}{p_{k|l}=}
    (1-\mu)\,p_l + \mu, & \text{for} \quad $k=l$ \label{eq:8_2}\\
    (1-\mu)\,p_{k}, & \text{for} \quad $k \neq l$.\label{eq:18_2}
    \end{numcases}
\subsection{Unitary dilation}
Given the Kraus operators in Eqs.\ \eqref{eq:K_R1} and \eqref{eq:K_Rt}, a suitable environment can be set up for the (walker+system)'s open dynamics.\ The environment in our case turns out to be 2 two-level systems denoted `molecules' or \textit{ancillas} so that the (ancillas+walker+system) together evolves unitarily during the collisions.\ This process of constructing an unitary operator in an extended Hilbert space that reproduces the reduced dynamics of the system of interest, is referred to in the literature as \textit{unitary dilation} \cite{N&C,Preskill,vom_Ende}.\\
If we consider the ancilla systems to be in the pure state $\rho_e = |00\rangle\langle00|$ at the beginning of each collision with the (system+walker), the unitary collision operator is constructed so that the Kraus operators are given as 
\begin{equation}
    K_{\alpha} = \langle \alpha |U|00\rangle
    \label{eq:kraus_uni}
\end{equation}
where $\alpha$ corresponds to the ancilla states $\{|00\rangle,|01\rangle,|10\rangle,|11\rangle\}$.\ This ensures, as described in Eqs.\ \eqref{eq:sep_rho} and \eqref{eq:Op_sum}, that 
    $R_t =$ $\text{Tr}_{e}\{U(R_{t-1}\otimes \rho_e)U^\dagger\} =$ $\sum_{\alpha=1}^{4}\ K_{\alpha}\ R_{t-1}\  K_{\alpha}^{\dagger}$.

The ($2^{n+3}\times 2^{n+3}$) unitary operator $U_0$ governing the collision process corresponding to \eqref{eq:K_R1} and satisfying \eqref{eq:kraus_uni} has been constructed as
\begin{adjustwidth}{-.5cm}{-.7cm}
  \begin{equation}
  \begingroup 
\setlength\arraycolsep{-2pt}
  U_0 = \begin{psmallmatrix} 
    \sqrt{p_{g}}G &0&0&0&0 &\sqrt{p_{g'}}G &0 &0\\ 
    0&0&\sqrt{p_{g}}G &0&0 &0 &0&\sqrt{p_{g'}}G \\
    0&\sqrt{p_{g}}G &0&0&\sqrt{p_{g'}}G&0 &0 &0 \\
    0&0&0&\sqrt{p_{g'}}G &0 &0&\sqrt{p_{g}}G &0 \\
    0&0&\sqrt{p_{g'}}G' &0&0 &0 &0&-\sqrt{p_{g}}G'\\
    \sqrt{p_{g'}}G' &0&0&0&0 &-\sqrt{p_{g}}G' &0 &0\\ 
    0&0&0&\sqrt{p_{g}}G' &0 &0&-\sqrt{p_{g'}}G' &0 \\
    0&\sqrt{p_{g'}}G' &0&0&-\sqrt{p_{g}}G'&0 &0 &0 
    \end{psmallmatrix}
    \endgroup
    \label{eq:Unitary_0}
\end{equation}
\end{adjustwidth}

The unitary $U$ corresponding to \eqref{eq:K_Rt}, i.e., for all times $t\geq 2$, can be similarly found to be
\begin{adjustwidth}{-.7cm}{-.7cm}
  \begin{equation}
  \begingroup 
\setlength\arraycolsep{-2pt}
  U = \begin{psmallmatrix} 
    \sqrt{p_{g|g}}G &0&0&0&0 &\sqrt{p_{g'|g}}G &0 &0\\ 
    0&0&\sqrt{p_{g|g'}}G &0&0 &0 &0&\sqrt{p_{g'|g'}}G \\
    0&\sqrt{p_{g|g'}}G &0&0&\sqrt{p_{g'|g'}}G&0 &0 &0 \\
    0&0&0&\sqrt{p_{g'|g}}G &0 &0&\sqrt{p_{g|g}}G &0 \\
    0&0&\sqrt{p_{g'|g'}}G' &0&0 &0 &0&-\sqrt{p_{g|g'}}G'\\
    \sqrt{p_{g'|g}}G' &0&0&0&0 &-\sqrt{p_{g|g}}G' &0 &0\\ 
    0&0&0&\sqrt{p_{g|g}}G' &0 &0&-\sqrt{p_{g'|g}}G' &0 \\
    0&\sqrt{p_{g'|g'}}G' &0&0&-\sqrt{p_{g|g'}}G'&0 &0 &0 
    \end{psmallmatrix}
    \endgroup
    \label{eq:Unitary_t}
\end{equation}
\end{adjustwidth}
We can decompose the unitary $U$ into parts that act on different positions of the total (ancillas+walker+system).\ This is shown in Fig.\ \ref{fig:U_decomp}.\ We find that, $U = (\text{controlled}-\bigchi)_{e_1 s}\ (M)_{e_1 e_2 w}\ (G)_{s}$, where $\bigchi G = G'$ as was used in the previous chapter.\ The explicit expression for unitary $M$ is also given in the figure, with $A$ and $B$ being two $4\times 4$ matrices:
\begin{equation}
    A = \begin{psmallmatrix} 
    0 & \frac{\sqrt{p_{g|g}}-1}{\sqrt{p_{g'|g}}} & 0 & 0\\
    \frac{\sqrt{p_{g|g'}}}{\sqrt{p_{g'|g'}}}  & 0 & 0 & \frac{-1}{\sqrt{p_{g'|g'}}}\\
   \frac{-1}{\sqrt{p_{g'|g'}}}  & 0 & 0 & \frac{\sqrt{p_{g|g'}}}{\sqrt{p_{g'|g'}}}\\
   0 & 0 & \frac{-\sqrt{p_{g|g}}}{\sqrt{p_{g'|g}}} & 0
    \end{psmallmatrix},\quad \text{and}, \quad
    B = \begin{psmallmatrix} 
    0 & \frac{\sqrt{p_{g|g}}+1}{\sqrt{p_{g'|g}}} & 0 & 0\\
    \frac{\sqrt{p_{g|g'}}}{\sqrt{p_{g'|g'}}}  & 0 & 0 & \frac{1}{\sqrt{p_{g'|g'}}}\\
   \frac{1}{\sqrt{p_{g'|g'}}}  & 0 & 0 & \frac{\sqrt{p_{g|g'}}}{\sqrt{p_{g'|g'}}}\\
   0 & 0 & \frac{\sqrt{p_{g'|g}}+1}{\sqrt{p_{g|g}}} & 0
    \end{psmallmatrix}
\end{equation}
\subsubsection{\textit{Origin of `memory' in the noise}}
\begin{figure}[h!]
    \centering
    \captionsetup{width=\linewidth}
    \includegraphics[width=.99\linewidth]{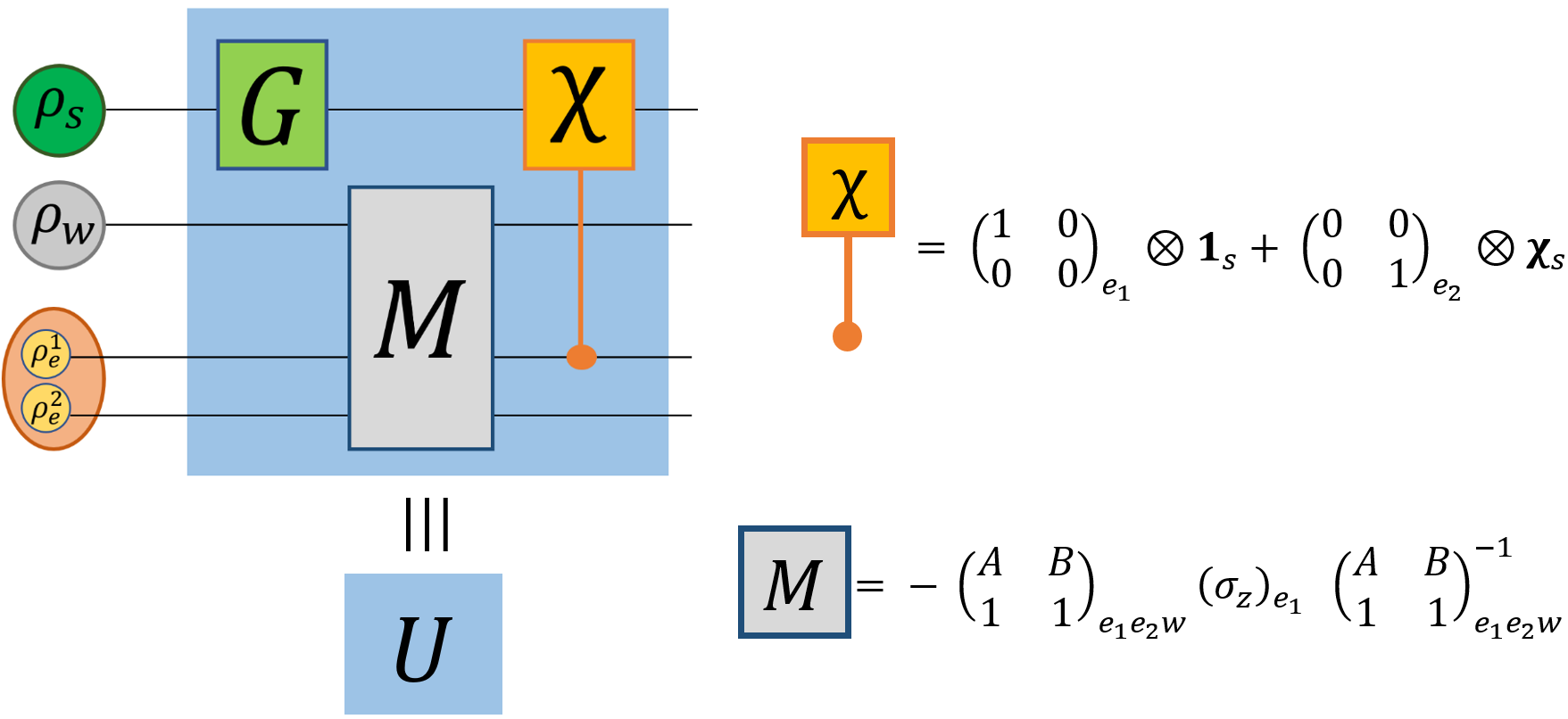}
    \caption[Decomposition of collision unitary]{The unitary $U$ is shown in an equivalent three-fold decomposition that act on different locations in (ancillas+walker+system).\ The ancillas are denoted here as $\rho_e ^1$ and $\rho_e ^2$.\ The explicit forms of $M$ (gray) and controlled-$\chi$ (orange) are shown in the right.\ }
    \label{fig:U_decomp}
\end{figure}

From the decomposition in Fig.\ \ref{fig:U_decomp}, the \textit{origin of `memory'} in the noise can be more clearly understood.\ At each time step, the total unitary on (ancillas+walker+system) is $U$.\ In a given time step, the system (register) first evolves under the noiseless Grover unitary $G$.\ Then, the unitary $M$ acts on (ancillas+walker); thus transforming the state $\rho_e ^1$ of the first ancilla qubit such that it then probabilistically controls the noisy $\bigchi$ evolution of the system.\\ Although the two ancillary qubits $\rho_e$ are `refreshed' at each time step, the memory of the system's noisy evolution $\bigchi$ is then passed on to the subsequent time step's $M$-evolution by the information in the walker's state $\rho_w$.\ This information is used in the next time step's $M$-evolution to evolve the state of $\rho_e ^1$ so that the controlled-$\bigchi$ in that time step is applied on the system according to the conditional probabilities $p_{i|j}$ in Eqs.\ \eqref{eq:8_2} and \eqref{eq:18_2}.
\begin{figure}[h!]
    \centering
    \captionsetup{width=\linewidth}
    \includegraphics[width=.5\linewidth]{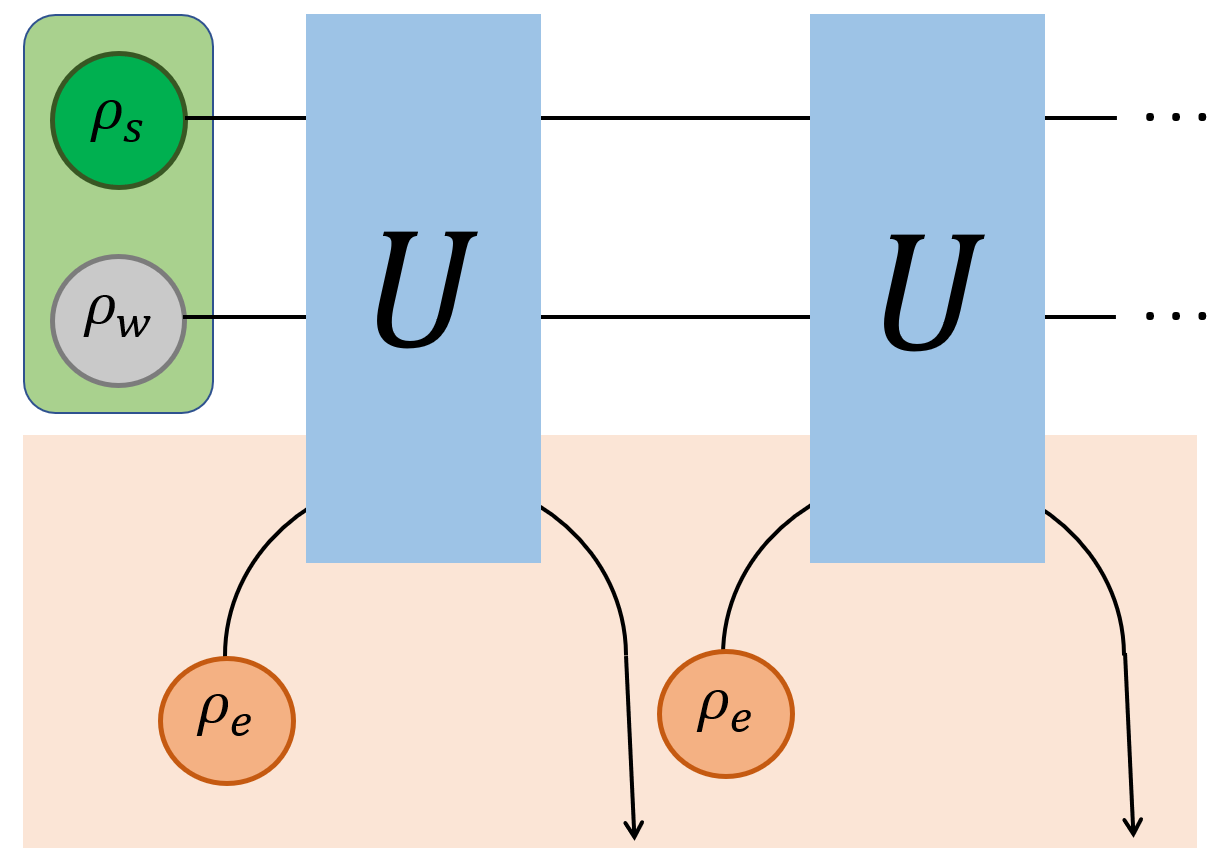}
    \caption[Collisional model for the (system + walker) time evolution]{Collisional model for the (system + walker) time evolution, Eq.\ \eqref{eq:K_R1}, \eqref{eq:K_Rt}.\ At each time step, the `ancillas' (composed of two 2-level systems) with state $\rho_e$ collides with the states $\rho_w$ and $\rho_s$.\ After each collision described by the unitary $U$, the ancillas are discarded.\ Memory effects in $\rho_s$ evolution is due to the coupling with $\rho_w$ through-out the evolution.}
    \label{fig:coll_model}
\end{figure}
\subsection{Chain of collisions}
Thus, as opposed to the application of the Grover unitary $G$ at each time step in case of the ideal Grover algorithm, the algorithm under Markovian-correlated noise (Section \ref{Example}) can instead be modeled as evolving under collisions with a walker and two `fresh' ancillas at each time step.\\
Since the time-evolution of the (system+walker) can be expressed in Kraus representation and that it has no initial correlations with the ancillas, imply that the dynamics is CPTP.\ Also, since fresh ancillas collide and evolve the (system+walker) state $R_t$ at each time step, implies the dynamics of $R_t$ is also CP-divisible and given by a \textit{discrete dynamical semigroup}
\begin{equation}
    \Phi(t,t_0)=\Phi(t-s,t_0)\circ \Phi(s,t_0); \quad t\geq s\geq t_0
\end{equation}
similar to the homogeneous composition law in \eqref{eq:semigroup} for Markovian evolution.\ Thus from the statement \ref{CP_marko}, the (system+walker)'s discrete-time dynamics can be treated as \textit{Markovian} \cite{RHP,Breuer_NM_rev}.

\section{Non-Markovian system evolution}
Although the (system+walker) state $R_t$ undergoes Markovian evolution, the reduced dynamics of the system's state $\rho_s=\text{Tr}_w \{R_t\}$ alone is non-Markovian, in general \cite{Jiang}.\ Since the walker is correlated at all times with the system, back and forth exchange of information is possible between them.\ The backflow of information from the walker into the system can then be quantified by the \textit{BLP}-measure of non-Markovianity $\mathcal{N}_{BLP}$ in Eq.\ \eqref{eq:N_BLP}.\ To reiterate, for discrete-time as in our case,
\begin{equation}
    \mathcal{N}_{BLP}(\Phi(t,t_0)) = \underset{\rho^1(t_0),\rho^2(t_0)}{\mathrm{max}} 
    \sum_{ \Delta\, D >0} \Delta\,  D(\rho^1(t),\rho^2(t))
    \label{eq:N_BLP_1}
\end{equation}
where $\Delta\,  D(\rho^1(t),\rho^2(t))$ is the change in trace distance between the time-evolved system states from time $(t-1)$ to $t$, with initial states $\rho^1(0)$ and $\rho^2(0)$.

\subsection{Quantifying information flow into system}
\label{Sec_quant_info}
\begin{figure}[h!]
\centering
    \includegraphics[width=0.86\linewidth]{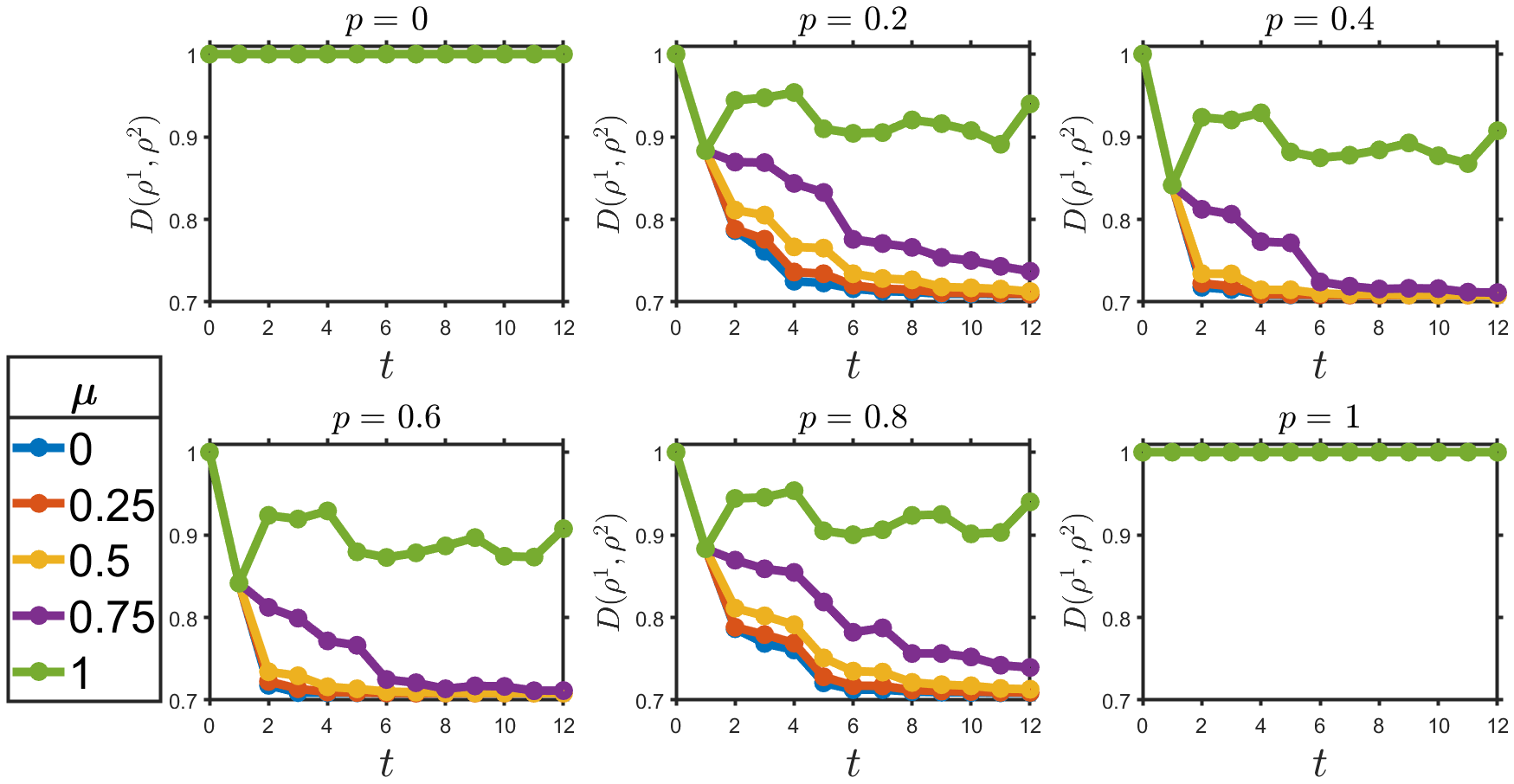}
    \caption[Trace distance as a function of time]{Trace distance $D(\rho^1(t),\rho^2(t))$ is plotted as a function of time $t$.\ Here $\rho^1(0)=|s\rangle\langle s|$ and $\rho^2(0)$ is as in Eq.\ \eqref{eq:rho_2}.\ The subplots are for different noise probabilities $p$ and the colored curves are for different values of the memory parameter $\mu$.\ For small values of $\mu$, $D(\rho^1,\rho^2)$ decreases monotonically, whereas for $\mu = 0.75$ (say), it is not monotonic and indicates `backflow' of information.\ The BLP-measure of non-Markovianity Eq.\ \eqref{eq:N_BLP_1} is then used to make the plots in Fig.\ \ref{fig:NM_mu}.}
\end{figure}
As was shown in Section \ref{sec:N_BLP}, $\mathcal{N}_{BLP}>0$ implies there is information flow from the environment to the system.\ The measure, Eq.\ \eqref{eq:N_BLP_1}, also requires maximizing over all system initial states.\ Although the maximization seems hard to compute directly, there exist pairs of initial states, called the \textit{optimal state pairs}, that achieve this maximum.\ It has been shown in \cite{weissman} that the optimal state pairs need to be orthogonal\footnote{If the two eigenspaces containing eigenvectors with non-zero eigenvalues of any two density matrices $\rho^1$ and $\rho^2$ are orthogonal, then $\rho^1$ is defined to be \textbf{\textit{orthogonal}} to $\rho^2$; denoted as $\rho^1 \perp \rho^2$.} and to lie on the boundary\footnote{$\rho$ is an \textit{interior point} of the state space $S(\mathcal{H})$ \textit{iff} $\forall \sigma \in S(\mathcal{H}) $, $\exists \lambda > 1$ so that $((1-\lambda)\sigma + \lambda \rho) \in S(\mathcal{H})$. If $\rho$ is not an interior point, then it is defined to be on the \textbf{\textit{boundary}} $\partial S(\mathcal{H})$ of $S(\mathcal{H})$. If $\rho$ has a zero eigenvalue, then $\rho \in \partial S(\mathcal{H})$.}.\\
In our case, we will take one of the system initial states in Eq.\ \eqref{eq:N_BLP_1} to be $\rho^1(0) = |s\rangle\langle s|$, where $|s\rangle$ is the uniform superposition of all system (register) states, see Eq.\ \eqref{eq:unif_sup_}.\ Note that $\rho^1(0)$ has only one non-zero eigenvalue and the zero eigenvalue is $(N-1)$-degenerate, implying that $\rho^1(0)$ belongs to the boundary of the state space.\ Moreover, the single `eigenstate with non-zero eigenvalue' of $\rho^1(0)$ happens to be orthogonal to the only `eigenstate with non-zero eigenvalue' of another state of the system, 
\begin{equation}
    \rho^2(0) = \frac{1}{N} \begin{psmallmatrix} \mathds{1}_{\frac{N}{2}} & -\mathds{1}_{\frac{N}{2}}\\ -\mathds{1}_{\frac{N}{2}} & \mathds{1}_{\frac{N}{2}}
    \end{psmallmatrix},
    \label{eq:rho_2}
\end{equation}
which will be used in the measure, Eq.\ \eqref{eq:N_BLP_1}.\\

 \begin{wrapfigure}[21]{l}{0.48\textwidth}
\centering
    \includegraphics[width=0.99\linewidth]{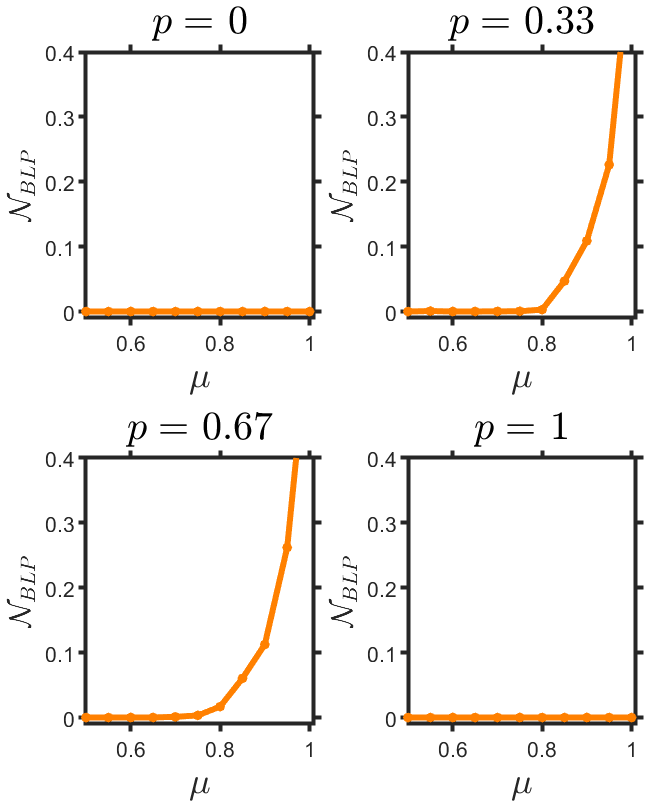}
    \caption[$\mathcal{N}_{BLP}$ is plotted with $\mu$, for different noise probabilities $p$]{$\mathcal{N}_{BLP}$ is plotted with respect to $\mu$, for different values of noise probability $p$. For $p=0$ and 1, $\mathcal{N}_{BLP}=0,\ \forall \mu$.\ In case of $p=0.33$ and 0.67, $\mathcal{N}_{BLP}>0$ for $\mu\gtrsim0.7$.\ Plots are for $n=3$, $\sigma_x$ noise and evolution up to $t=45$. }
    \label{fig:NM_mu}
\end{wrapfigure}For $p_{g'}=p=0$, i.e., \textit{noiseless} case, the system evolves under the ideal Grover unitary $G$.\ In this case, there is no information exchange between the system and the walker.\ Thus, $\mathcal{N}_{BLP}=0$ in case of $p=0$ for all parameters.\\
For $p_{g'}=p=1$, i.e., \textit{completely noisy} case, we have $p_{g|g} = \mu$, $p_{g'|g} = (1-\mu)$, $p_{g|g'} =0$, $p_{g'|g'} = 1$ from Eq.\ \eqref{eq:8_1}.\ In this case, the same $G'$ (the noisy Grover unitary) is applied at all times.\ Thus, in case of $p=1$ also, $\mathcal{N}_{BLP}=0$ for all parameters.\\
For $0<p<1$, $\mathcal{N}_{BLP}$ becomes non-zero only after the memory parameter $\mu$ becomes greater than a certain threshold value, as shown in Figure \ref{fig:NM_mu}.\  
This implies, for intermediate values of $p$ and sufficiently high $\mu$, backflow of information occurs
from the environment into the system.\ Also, it is interesting to point out that although such backflow is not seen for low $\mu$'s in case of $0<p<1$, it \textit{does not imply Markovianity}.

\subsection{CP-Divisibility and non-Markovianity}
\label{Sec_quant_CP}
The statement \ref{st_Cp_d} means that
\begin{gather}
    \mbox{\textit{if} there is an initial state $R$ on $\mathcal{H}_{S'}\otimes\mathcal{H}_S$} \nonumber\\ \mbox{so that the trace norm $||\big(\mathds{1}_{S'}\otimes (\Phi_t)_{S}\big)[R]||_1$ \textit{increases} temporarily,} \nonumber \\ \mbox{\textit{then} the discrete-time dynamics $\{\Phi_t\}_{t\geq0}$ is \textit{not} CP-divisible.}
    \label{impl_Cp}
\end{gather}

To show the above, we will take the initial state as $R=\frac{\mathds{1}_N}{N}\otimes (|s\rangle\langle s|-|w\rangle\langle w|)$, where $|s\rangle$ is the uniform superposition of all states of the register, $S$ and $|w\rangle$ is the marked pure state of the register.\ $S'$ is taken to be of the same dimension as $S$ and it is in the completely mixed initial state $\frac{\mathds{1}_N}{N}$.\\
We can define a \textit{rough} (in the sense that it is still dependent on the state $R$) measure that gives an idea of the CP-divisibility of this process.\ 
\begin{equation}
    \mathcal{N}_{CP} = \sum_{\mathclap{\substack{t\\ \Gamma_{t+1} >\Gamma_t}}}\ \big(\Gamma_{t+1}-\Gamma_t\big)
    \label{eq:N_cp_ga}
\end{equation}
where $\Gamma_t =\frac{1}{2}||\big(\mathds{1}_{S'}\otimes (\Phi_t)_{S}\big)[R]||_1$.\ Note that although this looks similar to Eq.\ \eqref{eq:N_BLP}, there is no maximization being performed (although ideally it should be, to be a universal measure).\ We are considering the trace distance in the $\mathcal{H}_{S'}\otimes\mathcal{H}_S$ space whereas $\mathcal{N}_{BLP}$ considers the trace distance in only $\mathcal{H}_S$ space.\ It is very important to note here, that although any \textit{non-zero} value of $\mathcal{N}_{CP}$ implies \textit{breaking} of CP-divisibility from statement \eqref{impl_Cp}, $\mathcal{N}_{CP}=0$ \textit{does not} necessarily imply that the process is \textit{CP-divisible}. \\
Here we are not using the $\mathcal{N}_{RHP}$ measure (Eq.\ \eqref{eq:N_RHP}) that was defined for continuous time evolution because the time evolution in our case is  discrete.\ But, $\mathcal{N}_{CP}$ is very similar to $\mathcal{N}_{RHP}$ and thus can be thought of as the discrete-time analogue of $\mathcal{N}_{RHP}$.\\

\begin{figure}[h!]
\centering
    \includegraphics[width=.66\linewidth]{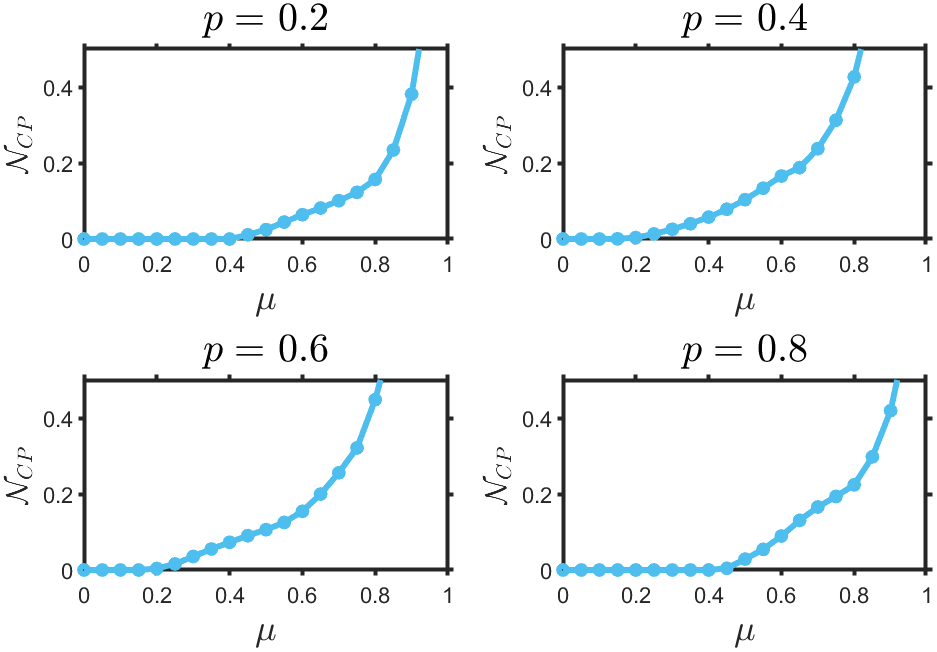}
    \caption[$\mathcal{N}_{CP}$ is plotted with $\mu$ for different $p$'s]{$\mathcal{N}_{CP}$ is plotted with $\mu$ for different values of $p$.\ For $0<p<1$, we see that $\mathcal{N}_{CP}>0$ for $\mu$ above some threshold - implying breaking of CP-divisibility in those cases.\ Here $n=3$, with $\sigma_x$ noise and time evolution up to $t=20$.}
    \label{fig:N_cp}
\end{figure}
As previously discussed, in case of $p=0$ (completely noiseless) and $p=1$ (completely noisy), the processes are unitary and CP-divisibility does not make sense in these cases.\ So, we have not plotted these two cases in Figure \ref{fig:N_cp}.\ From previous section, we have also found that there is no information back-flow in these cases too.\\
In our model, we find that for $0<p<1$, the  $\mathcal{N}_{CP}$ measure is non-zero for some values of $\mu$ above a threshold that depends on the values of $p$ and $n$.\ When $\mathcal{N}_{CP}>0$ in these cases, it means that for those parameters, the discrete-time dynamics of the register is guaranteed to break CP-divisibility and thus \textit{are non-Markovian}.\\ But, we also see in Figure \ref{fig:N_cp}, that for some small values of $\mu$, the measure $\mathcal{N}_{CP}=0$.\ In these particular cases, it is \textit{incorrect to claim} that these processes are necessarily CP-divisible.\ Since our measure Eq.\ \eqref{eq:N_cp_ga} is dependent on the initial state $R$, we can only say that in these cases the measure can not detect if they break CP-divisibility or not.\\
The \textit{main message being attempted to convey} through the analysis in Sections \ref{Sec_quant_info} and \ref{Sec_quant_CP} is that we can contrast the range of parameters in which each of the non-Markovianity measures (information flow and CP-divisibility) quantify the given processes as non-Markovian.\ We notice, by comparing Figures \ref{fig:NM_mu} and \ref{fig:N_cp}, that the measure quantifying CP non-divisibility identifies a greater portion of the parameter space than that identified by the measure quantifying information back-flow as non-Markovian.\ This is in accordance with the claim made in Section \ref{sec:N_BLP}, that $\mathcal{N}_{BLP}$ is a stronger measure of non-Markovianity (in the continuous-time evolution of a system, quantifies breaking of P-divisibility of the dynamical map) than the $\mathcal{N}_{RHP}$ measure (in continuous-time case, quantifies breaking of CP-divisibility of the map).\ 

\section{Effect of an elementary thermal bath}
Instead of taking the ancillas to be in a pure state $\rho_e = |00\rangle\langle00| = \begin{psmallmatrix} 1 & 0\\ 0 & 0\end{psmallmatrix}\otimes \begin{psmallmatrix} 1 & 0\\ 0 & 0\end{psmallmatrix}$ as we did in Eq.\ \eqref{eq:kraus_uni}, we can consider them all to be initially in thermal state 
\begin{equation}
    \rho_e = \begin{psmallmatrix} z_1 & 0\\ 0 & z_2\end{psmallmatrix}\otimes \begin{psmallmatrix} z_1 & 0\\ 0 & z_2\end{psmallmatrix} \label{eq:thermal_ancilla}
\end{equation}
where $z_1=\frac{1}{1+e^{-1/T}}$, $z_2=\frac{e^{-1/T}}{1+e^{-1/T}}$ with $T$ denoting a \textit{dimensionless temperature} parameter.\ Since these ancillas collide with the (system + walker) at every time step with the same initial thermal state, they mimic a \textit{thermal bath} \cite{ciccarello}.\ Whereas an actual thermal bath is made up of a continuum of modes, in this model, the interaction happens only with small, discrete and identically prepared constituents of the \textit{elementary} thermal bath. \\
\begin{wrapfigure}[20]{l}{0.48\textwidth}
    \centering
    \includegraphics[width=\linewidth]{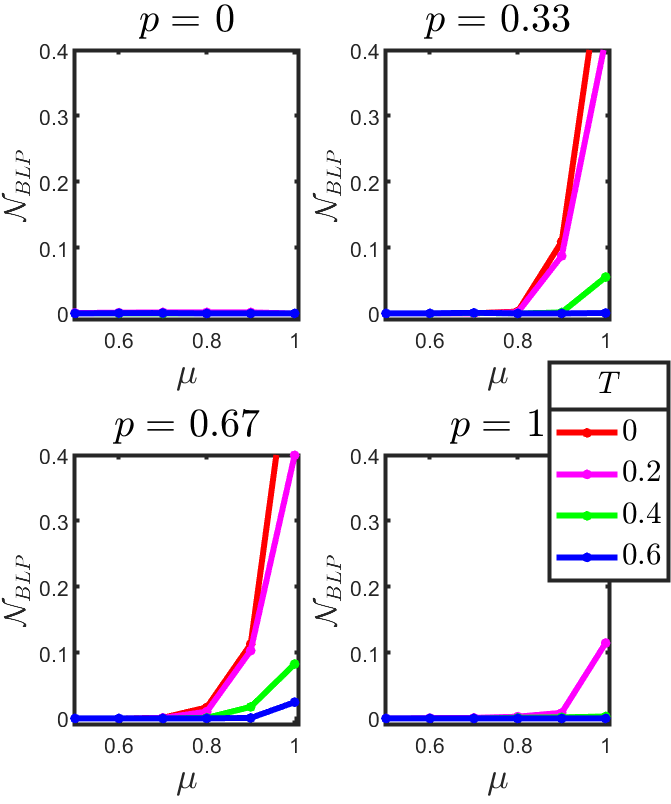}
    \caption[$\mathcal{N}_{BLP}$ is plotted with $\mu$, $p$ and Temperature $T$]{$\mathcal{N}_{BLP}$ is plotted with $\mu$, for different $p$'s.\ Colored curves in each subplot are for temperatures $T$ (inset) of the ancillas.\ Here, $n=3$, noise unitary $\sigma_x$ and the system evolution is up to $t=45$.}
    \label{fig:NM_mu_T}
\end{wrapfigure} Keeping the unitary $U$ describing the (ancillas + walker + system) collision same as in Eq.\ \eqref{eq:Unitary_0} and \eqref{eq:Unitary_t} and using the initial state of the ancillas at each collision to be $\rho_e$ in Eq.\ \eqref{eq:thermal_ancilla}, we can write the time evolution of the (walker + system) state at each time step in the following Kraus representation
\begin{equation}
    R_t = \sum_{\mathclap{\substack{\alpha,\beta \\\in \{00,01,10,11\}}}}\ K_{\alpha\beta}\ R_{t-1}\  K_{\alpha\beta}^{\dagger} \label{eq:K_Rt_1}
\end{equation}
\begin{equation}
K_{\alpha \beta} = \pi_{\beta}\  \langle \alpha |U|\beta\rangle;\end{equation}\begin{equation}\pi_{\beta}=
\begin{cases}
  z_1, & \text{for} \quad \beta =00,\\
  \sqrt{z_1 z_2}, & \text{for} \quad \beta \in \{01,10\},\\
  z_2, & \text{for} \quad \beta =11,
\end{cases}
\end{equation}
Thus, we can simulate coupling to a heat bath by making the (system + walker) collide with the thermal ancillas at every time step of the algorithm.\ 
Since a Kraus decomposition exists, the (system + walker) evolution is thus CP-divisible and as a result, Markovian, from statement \ref{CP_marko}.\ It turns out that the process is also non-unital and leads to energy dissipation to the bath.

\begin{figure}[h!]
\centering
    \includegraphics[width=.99\linewidth]{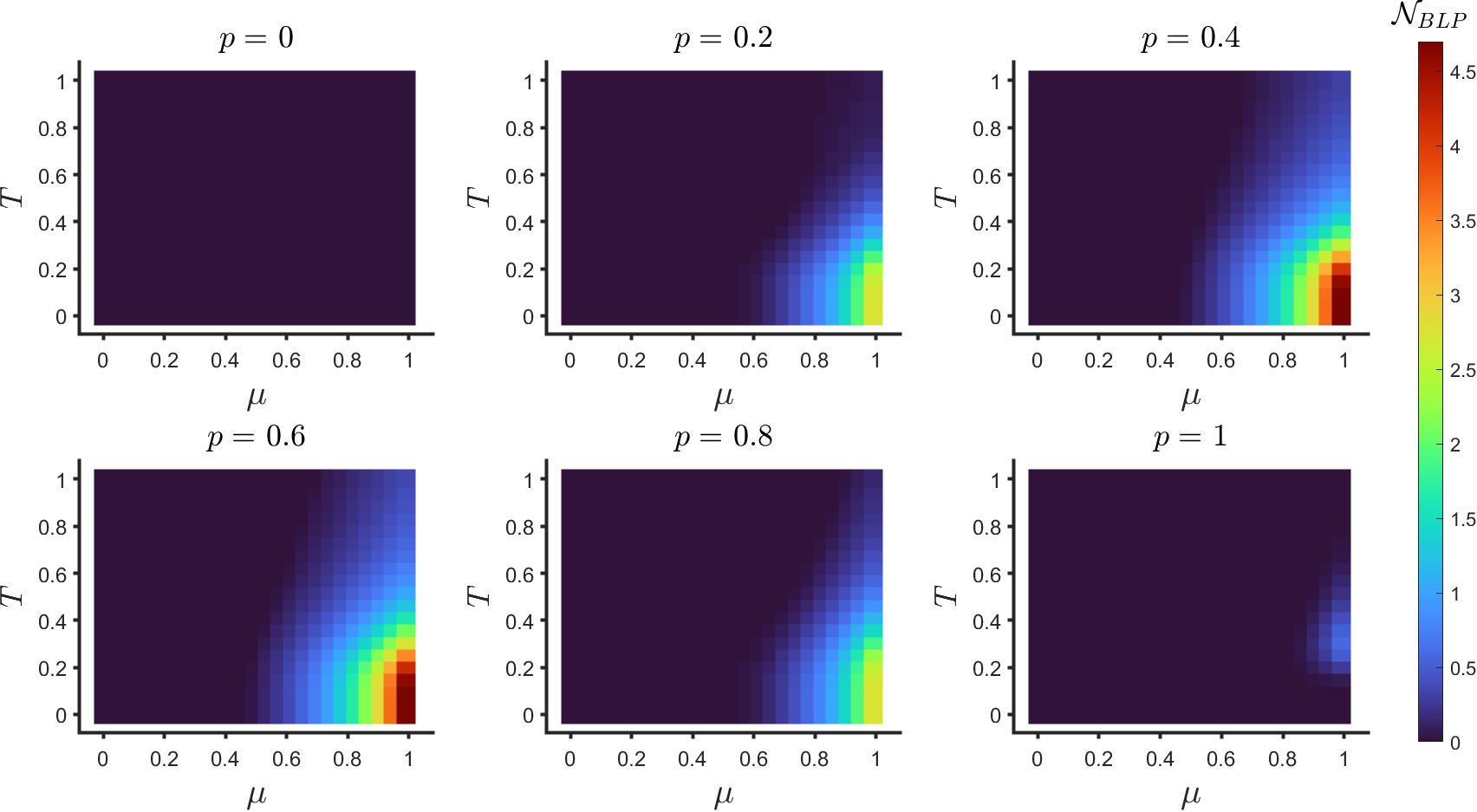}
    \caption[Variation of $\mathcal{N}_{BLP}$ with $T$ and $\mu$, for different $p$'s]{Variation of $\mathcal{N}_{BLP}$ (color-map on the right) with temperature $T$ and $\mu$ for different values of $p$.\ $\mathcal{N}_{BLP}$ increases with increasing $\mu$ and decreasing $T$.\ Here $n=3$, noise unitary is $\sigma_y$ and evolution up to $t=30$.}
    \label{fig:2d_NM}
\end{figure}

Figure \ref{fig:NM_mu_T} indicates the parameter ranges in which there is information back-flow into the system from the environment.\\ We have used the measure Eq.\ \eqref{eq:N_BLP_1} with the initial states $\rho^1(0)=|s\rangle\langle s|$ and $\rho^2(0)$ in Eq.\ \eqref{eq:rho_2}.\ In the completely noiseless case, i.e., $p=0$, we have $\mathcal{N}_{BLP}\approx 0,\ \forall \mu, T$.\ But for $p>0$, the effect of temperature on the information back-flow is more apparent. We notice that for increasing temperature, $\mathcal{N}_{BLP}$ decreases at a given $\mu$ and $p$, indicating increase in net outflow of information from (system + walker) to the bath.\\ 
While comparing Figure \ref{fig:NM_mu_T} with Figure \ref{fig:NM_mu}, it is also interesting to note that for $p=1$, $\mathcal{N}_{BLP}\neq 0$ for some values of $\mu$ and $T$.\ See Figure \ref{fig:2d_NM} for an overview of the effects of temperature, memory and noise probability on information back-flow.

We should note here that the measures $\mathcal{N}_{BLP}$ (Eq.\ \eqref{eq:N_BLP_1}) or $\mathcal{N}_{CP}$ (Eq.\ \eqref{eq:N_cp_ga}) directly depend on our consideration of the duration of the system's evolution.\ Since both these measures are based on trace distance increments, the value of the measure in general compounds over time, except when the trace distance is monotonically decreasing.\ Thus, although the values indicated in the figures above are not universal, a non-zero value of these measures is an obvious indicator of a non-Markovian process.\ Nevertheless, to help in comparisons, we have kept most of the other parameters equal, for example, the total number of qubits, the noise unitary and the total duration of evolution.

\vspace*{30mm}
\vspace{5cm}
\thispagestyle{empty}
\begin{center}
\begin{tikzpicture}
\node[align=center,draw,thick,minimum width=.8\textwidth,inner sep=9mm] (titlebox)%
{\fontshape{sc}\bfseries{\huge \fontsize{25}{35} Conclusions}};
\node[fill=white] (W) at (titlebox.north) {\bfseries \HUGE \color{gray}I I I};
\node (feat) at ([yshift=9mm]titlebox.north) {\textsc{\huge Part}};
\end{tikzpicture}
\end{center}
\addcontentsline{toc}{part}{Part-III: Conclusions}
\vspace{5cm}
\epigraph{``Ideas are like bundles of trajectories undergoing complicated evolution.''}{E.\ C.\ G. Sudarshan \textsuperscript
{\cite{sud}}\\ (1931-2018)\\\textit{Physicist}}
\bookmarksetup{startatroot}
\newpage
\pagestyle{empty}
{\color{graycolor}\rule[-1.5mm]{1cm}{.9cm}}\space \huge{\textbf{Summary}}\\
\vspace{-8mm}
\setstretch{.65}\par 
{\large 
In this thesis we have explored the situation when a quantum system is subjected to a noise that is correlated in time. Specifically, we consider that the noise is caused by free evolution of some of the qubits in a quantum register performing Grover's search.\ We show that irrespective of the presence of Markovian time-correlations in the noise, the necessary and sufficient condition for the algorithm's success probability to be unchanged while the number of noisy qubits is increased, is that the unitary operators governing the free evolution of the noisy qubits are the Pauli $\sigma_x$ and $\sigma_z$ matrices.\ For Pauli $\sigma_y$, success probability at all times will be unaltered as long as the number of noise sites is changed to another number with the same parity (even or odd).\\
We found that the system's noisy evolution can be modeled by colliding the system with three 2-level systems (qubits) at each time step.\ The system's non-Markovian dynamics can be thus made sense of as arising from the back-action of the walker qubit due to the system-walker coupling for all times, since otherwise the collisions with the two periodically-refreshed ancilla qubits lead to a Markovian evolution for the (system + walker).\ By decomposing the collision unitary, we could specifically point to the mechanism behind the Markovian-correlations of the noise in consecutive time steps.\ We then showed that for a sizeable portion of the parameter space, there is no `information back-flow' from the environment into the system.\ Still, in those cases, we found that the system can be classified as non-Markovian when the dynamical map is not CP-divisible.\\
Taking the ancilla qubits to be in their thermal states, we could simulate a thermal bath, albeit elementary.\ The effect of this was to decrease the information back-flow with increasing temperature of the bath.\\
\hfill\\
\textit{To conclude,} there are three main outcomes of this thesis.\ Firstly, we have shown that memory in noise can enhance the performance of the Grover algorithm under a Markovian-correlated noise.\ Secondly, we have shown that for a register's noisy qubits freely evolving under some unitary, the total time evolution of the register becomes independent of the number of noisy qubits if and only if the unitaries are the Pauli matrices.\ Thirdly, we found a physical model for the noisy register's environment and using that model, could explain the origin of memory in our particular case.
}

\vspace{5cm}
\par
\newpage
\pagestyle{empty}
{\color{graycolor}\rule[-1.5mm]{1cm}{.9cm}}\space \huge{\textbf{Acknowledgements}}\\
\vspace{-8mm}
\setstretch{.65}\par 
{\large
I want to thank Prof.\ Sen for giving me the opportunity to work on this project.\ His expert guidance and precious advice have been the main impetus behind bringing my efforts to fruition.\ I also learn and take inspiration from his openness, dedication and critique of each new direction of research that have been undertaken.\\
I want to thank Prof.\ Mahesh for his constant support and precious time as the local expert of the Thesis Advisory Committee.\ I am grateful to the physics department and all staff for their cooperation and the director of IISER Pune, Prof.\ Jayant B.\ Udgaonkar, for this wonderful opportunity.\\
The work done in this project would only have reached a fraction of its present volume, had Ahana and Chirag of Prof.\ Sen's group not offered me countless hours in discussions - pointing out mistakes, clearing doubts and providing suggestions during writing of this thesis.\ Your kindness has only raised my expectations from the people I will work with in the future.\ I have also been blessed with having some wonderful people around me.\ I wonder if some of my ideas in the project would have come to me, had my ruminating mind not been refreshed by the memorable times spent with Chrisil, Soubhadra, Hitesh, Vaibhav and others...\\
I acknowledge the support from the Department
of Science and Technology, Government of India through the INSPIRE scholarship that helped me sustain myself during the project.\\
Most of all, I thank my parents, my sister and other family members for their unconditional love and support during my studies.
}

\setstretch{1}


\appendix 

\sloppy
\printbibliography[heading=bibintoc]
\@openrighttrue\makeatother 

\end{document}